\documentclass[a4paper, notoc]{JHEP3}
\usepackage[dvips]{graphicx}
\usepackage{amsfonts}
\usepackage{amssymb}
\usepackage{epsfig}
\usepackage{cite}

\newcommand{\be}{\begin{equation}}
\newcommand{\ee}{\end{equation}}
\newcommand{\bea}{\begin{eqnarray}}
\newcommand{\eea}{\end{eqnarray}}
\newcommand{\mbb}{\mathbb}
\newcommand{\ti}{\times}
\newcommand{\half}{\frac{1}{2}}
\newcommand{\mc}{\mathcal}

\newcommand{\beqa}{\begin{eqnarray}}
\newcommand{\eeqa}{\end{eqnarray}}

 \newlength{\wth}
 \title{Mirror Mediation}
\author{Joseph P. Conlon \\ DAMTP,
  Centre for Mathematical
Sciences,
  Wilberforce Road, Cambridge, CB3 0WA, United Kingdom\\
 E-mail:
  \email{j.p.conlon@damtp.cam.ac.uk}
}

\abstract{
I show that the effective action of
string compactifications has a structure that can naturally solve the supersymmetric flavour and CP problems.
At leading order in the $g_s$ and $\alpha'$ expansions, the hidden sector factorises.
The moduli space splits into two mirror parts that depend on K\"ahler and complex structure moduli.
Holomorphy implies the flavour structure of the Yukawa couplings arises in only one part. In type IIA string theory
flavour arises through the K\"ahler moduli sector and in type IIB flavour arises through the complex structure moduli sector.
This factorisation gives a simple solution to the supersymmetric flavour and CP problems: flavour physics
is generated in one sector while supersymmetry is broken in the mirror sector. This mechanism does not require the presence of
gauge, gaugino or anomaly mediation and is explicitly realised by phenomenological models of IIB flux compactifications.
}

\preprint{DAMTP-2007-86}

\begin{document}

\section{Introduction}

TeV-scale supersymmetry is one of the most promising ideas for the new physics at the weak scale that will reveal itself at
the CERN Large Hadron Collider. This promise is because the presence of low-energy supersymmetry
cancels the quadratic divergences in the Higgs potential, stabilising it against radiative corrections.
Furthermore, supersymmetry gives a dynamical explanation for the structure of the Higgs potential through
the mechanism of radiative electroweak symmetry breaking, while low scale supersymmetry is also compatible with the absence of
large corrections to precision electroweak observables at LEP I.

Low energy supersymmetry is parametrised by the MSSM soft Lagrangian which specifies quantities such as squark
and gaugino masses as well as trilinear scalar A-terms. One of the most significant aspects of supersymmetric
phenomenology is that this Lagrangian has to take a very special form, due to the flavour and CP problems of
low-energy supersymmetry. These state that generic choices of the MSSM soft Lagrangian lead to large, new and unobserved
sources of flavour-changing neutral currents and CP violation. For example, if the squark masses of the first
two generations are not essentially degenerate, supersymmetric contributions to $K_0 - \bar{K}_0$ mixing
significantly exceed the observed rates. A full study of the flavour physics constraints on the MSSM spectrum can be found in
\cite{Bertolini1991, hepph9604387}, and for practical purposes these constraints can be summarised in the
requirements that
$$
(m^{I})^2_{\alpha \bar{\beta}} = (m^{I})^2 \delta_{\alpha \bar{\beta}}, \qquad A^I_{\alpha \beta \gamma} = A^I Y_{\alpha \beta \gamma}.
$$
That is, scalar masses of squarks and sleptons should be flavour-blind and the trilinear A-terms should be proportional to the Yukawa couplings.
These constraints hold within each set of gauge-charged fields - for example, $U_R$ and $D_R$ squarks need not have the same mass.

A satisfactory theoretical understanding of supersymmetry breaking requires an understanding
of why the soft terms should be flavour-universal. This
flavour problem has been the motivation for much of the work in supersymmetric model-building.
For example, gauge or gaugino mediation \cite{DFS81, DR81, hepph9801271, hepph9911293, hepph9911323} solves the flavour problem by
breaking supersymmetry at low energies and mediating to the observable sectors through flavour-blind
gauge interactions.
A different approach is that of anomaly mediation \cite{hepth9810155, hepph9810442},
where universal loop effects are used to generate soft masses, although due to tachyonic slepton masses
minimal anomaly mediation is not a viable phenomenological scenario.
Much of the motivation for models of gauge- or anomaly-mediated supersymmetry breaking lies in the assumption that
gravity-mediated supersymmetry breaking will automatically violate flavour
universality. The place to test this assumption is string theory, as the prime candidate for a theory of quantum gravity.

The study of supersymmetry breaking in string theory has gone through several stages. The original work was
carried out in the context of the heterotic string, with supersymmetry breaking and moduli stabilisation driven by
hidden sector gaugino condensation. Although full moduli stabilisation was not possible in that context,
supersymmetry breaking was analysed through a parametrisation of the goldstino direction. It was found that
soft terms were flavour universal for the case of dilaton-domination, where $F^S \neq 0$ while $F^T = F^U = 0$
\cite{Casas1990, hepth9204012, hepth9303040, hepph9308271, hepph9508258}.
In principle, the direction of the goldstino is a dynamical quantity determined by the moduli potential.
The study of the phenomenology of supersymmetry breaking has undergone a resurgence following the developments in
moduli stabilisation \cite{hepth0105097, hepth0301240, hepth0502058} (for reviews of moduli stabilisation see \cite{hepth0509003, hepth0610102, hepth0610327, hepth0701050}).
These allow the goldstino direction to be explicitly computed and the structure of
supersymmetric soft terms studied. This has been the subject of extensive research; recent papers studying this question
include
\cite{hepth0311241, hepth0406092, hepth0408036, hepth0410074, hepth0411066, hepth0411080,hepth0412150,
hepph0502151,hepth0503216,hepph0504036,hepph0504037,hepth0505076,
hepph0512081,hepth0605108,hepth0605141,hepth0605232,hepth0606262,hepph0610038,
hepth0610129,hepth0610297,hepth0611024,hepth0611279,hepph0612035,0701034, hepph0703024,hepph0703163,
hepph0702146, hepth0703105, 07040737, 07043403, 07071595, 07090221, 07094259}.
The purpose of the present paper is to provide a systematic
study of the necessary conditions for flavour universal soft terms and it
 extends arguments made in embryonic form in \cite{hepth0610129} (also see \cite{hepph9311352, hepth0611279}).

The structure of this paper is as follows. Section \ref{sec2} reviews the computation of
soft terms in supergravity and defines a set of sufficient conditions for the soft terms to be flavour-universal and CP-conserving.
These conditions are taken as the definition of mirror mediation. They correspond to the existence of a factorisation
of the hidden (moduli) fields into two sectors, with one sector generating the flavour structure and the other responsible for
supersymmetry breaking. At the level of effective field theory this structure is \emph{ad hoc}. Sections \ref{sec3} and
\ref{sec4} are devoted to showing that the mirror mediation structure is naturally realised within
type II string compactifications. In this case the susy-breaking and flavour sectors are associated with the
K\"ahler and complex structure moduli.
Section \ref{sec3} focuses on the factorisation of the moduli space and matter metrics, and
section \ref{sec4} on the structure of susy breaking that arises in IIB flux compactifications. An appendix studies
the properties of brane intersection angles in Calabi-Yau models of intersecting brane worlds; these enter into
the computation of soft scalar masses.

\section{Mirror Mediation}
\label{sec2}

The computation of gravity-mediated\footnote{As is well-known, the expression `gravity mediation'
is confusing as it does not mean `mediated by gravity'.
It refers instead to soft terms generated by non-renormalisable contact interactions in the supergravity Lagrangian.
This paper will follow convention and perpetuate this confusion.}
 soft terms follows a standard structure which it is useful to review.
$\mc{N}=1$ supergravity is specified at two derivatives by a K\"ahler potential, superpotential and gauge kinetic functions.
These are functions of the chiral superfields, which are separated into visible and hidden
sectors, $C^m$ and $\Phi_i$. $C^m$ denote the matter multiplets of the MSSM. These
are charged under the Standard Model gauge groups and giving them a vev reduces the rank of the gauge group.
$\Phi_i$ are the hidden sector fields (the moduli). These are uncharged and may have large vevs.
The full K\"ahler potential and superpotential can be
expanded in powers of the visible sector fields,
\bea
K & = & \hat{K}(\Phi_i, \bar{\Phi}_i) + \tilde{K}_{C_\alpha \bar{C}_\beta}(\Phi_i, \bar{\Phi}_i)
C^\alpha \bar{C}^\beta + (\tilde{Z}_{C_\alpha C_\beta}(\Phi_i, \bar{\Phi}_i) C^\alpha C^\beta + c.c) + \ldots, \\
W & = & \hat{W}(\Phi_i) + \mu_{\alpha \beta}(\Phi_i) C^{\alpha} C^{\beta} + Y_{\alpha \beta \gamma}(\Phi_i) C_{\alpha} C_{\beta} C_{\gamma} + \ldots, \\
f & = & f(\Phi_i).
\eea
The supergravity scalar potential is
\be
\label{fullv}
V_F = e^{K} \left( K^{i \bar{j}} D_i W D_{\bar{j}} \bar{W} - 3 \vert W \vert^2 \right),
\ee
where $D_i W = \partial_i W + (\partial_i K) W$.
The vevs of the hidden sector moduli $\Phi_i$ are determined by the hidden sector scalar potential, which is\footnote{We set $M_P = 1$.}
\be
\label{hiddv}
V_F = e^{\hat{K}} \left( \hat{K}^{i \bar{j}} D_i \hat{W} D_{\bar{j}} \bar{\hat{W}} - 3 \vert \hat{W} \vert^2 \right).
\ee
The moduli F-terms are $F^i = e^{\hat{K}/2} \hat{K}^{i \bar{j}} D_{\bar{j}} \bar{W}$ and the gravitino mass
$m_{3/2} = e^{\hat{K}/2} \vert W \vert$. In terms of these,
\be
V_F = \hat{K}_{i \bar{j}} F^i \bar{F}^{\bar{j}} - 3 m_{3/2}^2.
\ee
The F-terms parametrise supersymmetry breaking and enter into all expressions for soft supersymmetry
breaking parameters.

In gravity mediation, the vacuum is a supersymmetry-breaking minimum of (\ref{hiddv})
and the principal source of supersymmetry breaking is the F-terms of the
hidden sector moduli. To match the observed
cosmological constant, it is necessary that the F-term potential vanish, with $V_F = 0$
at the minimum. Supersymmetry breaking generates soft masses for scalars and gauginos, as well as trilinear A-terms.
These arise from expanding the supergravity Lagrangian in powers of the matter fields.
In particular, soft scalar masses and trilinear terms come from expanding (\ref{fullv}) in powers of the matter fields $C^\alpha$.
The
resulting soft scalar Lagrangian is
\be
\mc{L}_{soft} = \tilde{K}_{\alpha \bar{\beta}} \partial_\mu C^{\alpha} \partial^\mu \bar{C}^{\beta} + m^2_{\alpha \bar{\beta}} C^{\alpha} \bar{C}^{\beta}
+ \frac{1}{6}\left(A'_{\alpha \beta \gamma} C^{\alpha} C^{\beta} C^{\gamma} + c.c \right),
\ee
 where the unnormalised soft terms are given by
\bea
\label{scalarmass}
\tilde{m}_{\alpha \bar{\beta}}^{2} & = &
(m_{3/2}^2 + V_0) \tilde{K}_{\alpha
  \bar{\beta}}  - \bar{F}^{\bar{m}} F^n \left( \partial_{\bar{m}} \partial_n \tilde{K}_{\alpha \bar{\beta}}
- (\partial_{\bar{m}} \tilde{K}_{\alpha \bar{\gamma}}) \tilde{K}^{\bar{\gamma} \delta} (\partial_n \tilde{K}_{\delta \bar{\beta}})
\right) \\
\label{aterm}
A'_{\alpha \beta \gamma} & = & e^{\hat{K}/2} F^m \Big[ \hat{K}_m Y_{\alpha \beta \gamma} +
\partial_m  Y_{\alpha \beta \gamma}
- \left( (\partial_m \tilde{K}_{\alpha \bar{\rho}})
\tilde{K}^{\bar{\rho} \delta} Y_{\delta \beta \gamma} + (\alpha
\leftrightarrow \beta) + (\alpha \leftrightarrow \gamma)
\right) \Big]
\eea
$V_0$ is the vacuum cosmological constant and will be set to zero.
The gaugino masses are given by
\be
M_a = F^m \frac{\partial_m f_a}{2 \hbox{Re}(f_a)}.
\ee
In the case of diagonal matter metrics the soft terms can be written as
\bea \label{SoftMassDiagFormula} m_\alpha^2  &
=  & (m_{3/2}^2 + V_0) - F^{\bar{m}}F^n \partial_{\bar{m}}
\partial_n \log \tilde{K}_\alpha. \\
\label{ATermFormula} A_{\alpha \beta \gamma} & = & F^m \left[
\hat{K}_m + \partial_m
  Y_{\alpha \beta \gamma} - \partial_m \log (\tilde{K}_\alpha \tilde{K}_\beta \tilde{K}_\gamma) \right].
\eea
When $\hat{K} = \hat{K}(\Phi + \bar{\Phi})$, (\ref{ATermFormula}) can be written in an instructive way
\be
\hat{A}_{\alpha \beta \gamma} = F^m \frac{\partial \hat{Y}_{\alpha \beta \gamma}}{\partial \hbox{Re}(\Phi_m)},
\ee
where $\hat{A}_{\alpha \beta \gamma}$ and $\hat{Y}_{\alpha \beta \gamma}$ are the physical (normalised) A-terms and Yukawa couplings.
In minimal flavour violation the scalar masses are flavour-universal and the trilinear terms have the same structure
as the CKM matrix.
This is equivalent to the requirement that the mass
term $\tilde{m}_{\alpha \bar{\beta}}^2$ is proportional to the kinetic terms $\tilde{K}_{\alpha \bar{\beta}}$, and the
unnormalised A-terms $A'_{\alpha \beta \gamma}$ are proportional to the superpotential Yukawa couplings $Y_{\alpha \beta \gamma}$.

It is clear that this is not the case for generic supergravity theories; for arbitrary $Y_{\alpha \beta \gamma}$ and
$\tilde{K}_{\alpha \bar{\beta}}$ both conditions are violated by equations (\ref{aterm}).
In order for the soft terms to be flavour-universal and CP-preserving, supersymmetry breaking must decouple from flavour physics.
I start by stating a set of sufficient conditions on the effective supergravity theory for gravity-mediated soft terms
to be flavour-universal. The expression `mirror mediation' will be used to refer to any theory satisfying these assumptions.

\begin{enumerate}
\item
The hidden sector fields factorise into two classes, $\Psi_i$ and $\chi_j$.
\item
The fields $\Psi_i$ and $\chi_j$ have decoupled kinetic terms, with $\Psi$ satisfying a reality assumption:
the K\"ahler potential is a direct sum
\be
\mc{K} = \mc{K}_1(\Psi + \bar{\Psi}) + \mc{K}_2(\chi, \bar{\chi}),
\ee
allowing the K\"ahler metric to be written in block-diagonal form
as

\be
\mc{K}_{i \bar{j}} = \left( \begin{array}{cc} \mc{K}_{\Psi \bar{\Psi}} & 0 \\ 0 & \mc{K}_{\chi \bar{\chi}} \end{array}
\right).
\ee

\item
The superpotential and specifically the superpotential Yukawa couplings depend only on the field
$\chi$, with no $\Psi$ dependence:
\be
Y_{\alpha \beta \gamma}(\Psi, \chi) = Y_{\alpha \beta \gamma}(\chi).
\ee
For the gauge kinetic functions the dependence is reversed: these depend linearly on the $\Psi$ fields, with no $\chi$ dependence,
\be
f_a(\Psi, \chi) = \sum_i \lambda_i \Psi_i.
\ee
\item
The matter metric factorises. For any set of fields $C^\alpha, C^{\beta}$ carrying the same gauge charges but of
different flavour, we can write
\be
\mc{K}_{\alpha \bar{\beta}}(\Psi, \bar{\Psi}, \chi, \bar{\chi}) = h(\Psi + \bar{\Psi}) k_{\alpha \bar{\beta}}(\chi, \bar{\chi})
\ee
with a universal dependence on $\Psi$. The function $h$ is allowed to vary
between fields of different gauge charges (e.g. between $U_R$
and $D_R$). This also implies the factorisation of the physical Yukawa couplings,
\be
\hat{Y}_{\alpha \beta \gamma}(\Psi, \bar{\Psi}, \chi, \bar{\chi}) =
\underbrace{\left(\frac{e^{\mc{K}_1/2}}{(h_1 h_2 h_3)^{\half}}(\Psi, \bar{\Psi})\right)}_{\textrm{$\Psi$-dependent prefactor}} \ti
 \underbrace{\left(e^{\mc{K}_2/2} (k^{\alpha \alpha'} k^{\beta \beta'} k^{\gamma \gamma'})^{\half}  Y_{\alpha' \beta' \gamma'}
 (\chi) \right)}_{\textrm{$\chi$-dependent flavour structure}}.
\ee
\item
The dynamics of the vacuum is such that the $\Psi$ fields are stabilised non-supersymmetrically and the
$\chi$ fields are stabilised supersymmetrically:
\be
D_{\Psi_i} W \neq 0, D_{\chi_j} W = 0.
\ee
Together with assumption 2, this is equivalent to the statement that $F^{\Psi} \neq 0, F^{\chi} = 0$.
\end{enumerate}

These assumptions define mirror mediation.
They construct two decoupled sectors $\Psi$ and $\chi$, such that the $\Psi$ sector breaks
supersymmetry and the $\chi$ sector generates the flavour structure.
It is easy to verify that these assumptions lead to soft terms that are flavour-universal and CP-preserving.
As only $F^{\Psi} \neq 0$ and as the matter metric factorises, we have
\bea
\tilde{m}_{\alpha \bar{\beta}}^{2} & = &
(m_{3/2}^2 + V_0) \tilde{K}_{\alpha
  \bar{\beta}} - \bar{F}^{\bar{\Psi}_j} F^{\Psi_i} \left( \partial_{\bar{\Psi}_j} \partial_{\Psi_i} \tilde{K}_{\alpha \bar{\beta}}
- (\partial_{\bar{\Psi}_j} \tilde{K}_{\alpha \bar{\gamma}}) \tilde{K}^{\bar{\gamma} \delta} (\partial_{\Psi_i} \tilde{K}_{\delta \bar{\beta}})
\right) \nonumber \\
& = & (m_{3/2}^2 + V_0) \tilde{K}_{\alpha
  \bar{\beta}} - \bar{F}^{\bar{\Psi}_j} F^{\Psi_i} \left( \partial_{\bar{\Psi}_j} \partial_{\Psi_i} h(\Psi, \bar{\Psi})
  - \frac{\partial_{\bar{\Psi}_j} h(\Psi, \bar{\Psi}) \partial_{\Psi_i} h(\Psi, \bar{\Psi})}{h(\Psi, \bar{\Psi})} \right)
  k_{\alpha \bar{\beta}}(\chi, \bar{\chi})
  \nonumber \\
  & = & \left( (m_{3/2}^2 + V_0)h - \bar{F}^{\bar{\Psi}_j} F^{\Psi_i} \left( \partial_{\bar{\Psi}_j} \partial_{\Psi_i} h
  - \frac{\partial_{\bar{\Psi}_j} h \partial_{\Psi_i} h}{h} \right) \right)(\Psi, \bar{\Psi}) k_{\alpha \bar{\beta}}(\chi, \bar{\chi})
\eea
As the mass squares are then a constant multiple of the kinetic terms, the soft masses are flavour diagonal. Now using
assumptions 3 and 4, we likewise obtain for the trilinear A-terms
\be
A_{\alpha \beta \gamma} = e^{\hat{K}/2} Y_{\alpha \beta \gamma}(\chi) \left( F^{\Psi} \partial_{\Psi} \hat{K}(\Psi, \bar{\Psi})
- 3 \frac{F^{\Psi} \partial_{\Psi} h(\Psi, \bar{\Psi})}{h(\Psi, \bar{\Psi})} \right),
\ee
which are manifestly proportional to the Yukawa couplings.

The reality condition on the K\"ahler metric for the $\Psi$ fields in assumption 2 is necessary to ensure that
there is no relative phase between different A-terms, with the phase of all A-terms being set by that of the goldstino F-term.
The linearity condition in assumption 3 likewise ensures that the gaugino mass phases are universal and aligned with those
of the A-terms. If a $\mu$-term is generated through the Giudice-Masiero mechanism \cite{GiudiceMasiero},
then the reality condition also ensures that
the phase of the $\mu$ term aligns with that of the A-terms and gaugino masses.

If a theory exhibits mirror mediation, the soft terms it generates are automatically flavour-universal and CP-preserving.
Formulated within effective field theory, the assumptions that enter mirror mediation
are \emph{ad hoc}: their entire purpose is to ensure flavour universality.
The purpose of this paper is to point out that in string theory the mirror mediation structure
is realised naturally and occurs in large classes of string compactifications.

\section{Mirror Mediation in String Theory}
\label{sec3}

We wish to relate the structure of mirror mediation to that appearing in string compactifications.
In string compactifications the hidden sector of effective field theory should be identified with the moduli of
the compactification. These are associated with the extra-dimensional geometry and enter into expressions for the gauge and
Yukawa couplings of the low-energy theory.
To preserve low-energy supersymmetry, string theory should be compactified on a Calabi-Yau manifold.
In this case the Calabi-Yau geometry naturally provides two main classes of moduli, K\"ahler and complex structure moduli.
The K\"ahler moduli describe the size of the Calabi-Yau and the complex structure moduli the shape.
In terms of the Hodge numbers of the Calabi-Yau there are $h^{1,1}$ K\"ahler moduli and $h^{2,1}$ complex structure moduli.
We will identify the two sectors $\Psi$ and $\chi$ required for mirror mediation with the K\"ahler and complex structure moduli.
Which way the identification goes will depend on whether the string theory involved is IIA or IIB.\footnote{The mirror mediation
structure relies on the existence of two separate classes of moduli, which at leading order decouple. In weakly coupled IIA, IIB and
heterotic models, the K\"ahler and complex structure moduli provide these two separate classes. There are however regimes in which
this breaks down. For example, in the case of M-theory, which is the strong coupling limit of IIA string theory,
there is only one class of moduli, which are associated with the geometry of 3-cycles.
In this case all moduli are on an equal footing and the mirror mediation structure is not possible.}

\subsection{Factorisation of Moduli Space}

The second assumption of mirror mediation is that the moduli space factorises.
In pure $\mc{N}=2$ type II string compactifications, this factorisation is exact and ensured by mirror symmetry.
In $\mc{N}=1$ type II orientifold compactifications, the factorisation still exists at leading order, being broken by
subleading corrections.
For IIA string theory, the classical K\"ahler potential (at leading order in the $g_s$ and $\alpha'$ expansions) is \cite{hepth0412277}
\be
\label{IIAk}
\mc{K} = - \ln (\mc{V}) - 2 \ln \left( \int \hbox{Re}\left(C \Omega \right) \wedge * \hbox{Re}\left(\bar{C \Omega}\right) \right) .
\ee
$C$ is a compensator field that incorporates the dilaton dependence.
For IIB string theory, we have
\be
\label{IIBk}
\mc{K}= - 2 \ln (\mc{V}) - \ln \left( i\int \Omega \wedge \bar{\Omega} \right) - \ln (S + \bar{S}).
\ee
In both cases the dependence of the K\"ahler potential on the K\"ahler moduli $T_i$ and the
 complex structure moduli $U_i$ factorises. As chiral superfields, the definitions of `K\"ahler moduli' and
`complex structure moduli' differ in IIA and IIB:
\bea
\textrm{IIB(D3/D7)}: & &  T = e^{-\phi} \hbox{Vol}(\Sigma_4) + i C_4, \qquad U = \int_{\Sigma_3} \Omega,  \\
\textrm{IIA}: & &  T = \hbox{Vol}(\Sigma_2) + i B_2,
\qquad U = e^{-\phi} \hbox{Vol}(\Sigma_3) + i C_3. \nonumber
\eea
$\Sigma_k$ refers to a cycle in the Calabi-Yau of dimensionality $k$ and the RR forms are understood to be integrated
over these cycles.
The different definitions of the moduli explain the apparent
different factors of the volume in (\ref{IIAk}) and (\ref{IIBk}).
In toroidal examples the K\"ahler potential has a simple expression for both IIA and IIB models,
\be
\mc{K} = - \ln(S + \bar{S}) - \ln \Big( (T_1 + \bar{T_1})(T_2 + \bar{T}_2)(T_3 + \bar{T}_3) \Big) -
 \ln \Big( (U_1 + \bar{U}_1)(U_2 + \bar{U}_2)(U_3 + \bar{U}_3) \Big)
\ee
From (\ref{IIAk}) and (\ref{IIBk})
we see that the K\"ahler potential admits a factorised form, consistent with
assumption 2. The K\"ahler metric can be written as
\be
\mc{K}_{i\bar{j}} = \left( \begin{array}{ccc} \mc{K}_{T\bar{T}} & 0 & 0 \\ 0 & \mc{K}_{U\bar{U}} & 0 \\
0 & 0 & \mc{K}_{S \bar{S}} \end{array} \right).
\ee
The different classes of moduli represent distinct sectors, with kinetic terms that do not mix with each other.

The factorisation is broken by subleading corrections, which lead to non-vanishing values for $\mc{K}_{T \bar{U}}$ and
$\mc{K}^{T \bar{U}}$. The discussion here will focus on the IIB case, but analogous results will hold in IIA.
One way factorisation can be broken is through the presence of D3 branes. For example, we can consider
the toroidal $T^6/\mbb{Z}_2 \ti \mbb{Z}_2$ orientifold in the presence of D3 branes. The K\"ahler potential is then
(e.g. see \cite{hepth0508171})
\be
\mc{K} = - \sum_{I=1}^3 \log \left[ (T_i + \bar{T}_i)(U_i + \bar{U}_i) + \frac{1}{8 \pi} (A_i + \bar{A}_i)^2 \right].
\ee
Here $A_i$ are brane moduli that parametrise the position of a D3 brane on the torus $i$.
The sum is over each of the three tori. It is straightforward to compute the resulting K\"ahler metric and its inverse.
The metric is a direct sum of three terms, one for each torus. Letting $I$ index the torus, we have
\be
\mc{K}_I^{-1} = \left(
\begin{array}{ccc}
(T_I + \bar{T}_I)^2 & - \frac{(A_I + \bar{A}_I)^2}{8 \pi} & (A_I + \bar{A}_I)(T_I + \bar{T}_I) \\
- \frac{(A_I + \bar{A}_I)^2}{8 \pi} & (U_I + \bar{U}_I)^2 & (A_I + \bar{A}_I)(U_I + \bar{U}_I) \\
(A_I + \bar{A}_I)(T_I + \bar{T}_I) & (A_I + \bar{A}_I)(U_I + \bar{U}_I) &
\left( \frac{(A_I + \bar{A}_I)^2 - 8 \pi (T_I + \bar{T}_I)(U_I + \bar{U}_I)}{2} \right)
\end{array} \right)
\ee
In the limit that the cycle sizes are large, the $T$ and $U$ sectors remain factorised. The figure of merit for this is
the ratio, $\eta$, of $(\mc{K}^{-1})^{T \bar{U}}$ and $(\mc{K}^{-1})^{T \bar{T}}$. These are given by
$$
\mc{K}^{T \bar{T}} = (T + \bar{T})^2, \qquad
\mc{K}^{U \bar{T}} = -\frac{(A + \bar{A})^2}{8 \pi}, \qquad
\eta = - \frac{(A + \bar{A})^2}{8 \pi (T + \bar{T})^2}.
$$
The reason why $\eta$ represents the figure of merit is that it gives
the cross-coupling induced from $F^T$ to $F^U$ given that
$D_T W \neq 0$ and $D_U W =0$. As
$$
F^T = e^{K/2} K^{T \bar{I}} D_{\bar{I}} W = e^{K/2} K^{T \bar{T}} D_{\bar{T}} W,
$$
$$
F^U = e^{K/2} K^{U \bar{I}} D_{\bar{I}} W = e^{K/2} K^{U \bar{T}} D_{\bar{T}} W,
$$
$\eta$ measures the ratio $F^U/F^T$, namely the
extent to which cross-couplings induce F-terms in the `wrong' sector.
This cross-coupling is suppressed by by a factor $(T + \bar{T})^{-2}$, and vanishes in the large-volume limit.

More generally, in the presence of a D3 brane with position moduli $\phi^i$ the $T$ fields are redefined as \cite{hepth0312232}
\be
\label{d3red}
T_{\alpha} = \hbox{Vol}(\Sigma_4) + i C_4 + (\omega_{\alpha})_{i \bar{j}} \phi^i \left( \bar{\phi}^{\bar{j}} - \frac{i}{2} \bar{z}^a (\chi_a)_l^{\bar{j}} \phi^l \right),
\ee
where $\omega_{\alpha}$ is the 2-form associated with $T_{\alpha}$ evaluated at the brane locus, and $\bar{z}^a$ and $\chi_a$
are the complex structure moduli and associated $(2,1)$ forms. For a single K\"ahler modulus model, the resulting K\"ahler potential
$\mc{K} = - 3 \ln (\hbox{Vol}(\Sigma_4))$ can be written as
\be
K = \mc{K}_{U \bar{U}} -3 \ln(T + \bar{T}) + \frac{\left(\phi \bar{\phi} + \phi \phi f(U, \bar{U})\right)}{T + \bar{T}} + \ldots
\ee
This gives
$$
K_{T \bar{T}} = \frac{3}{(T + \bar{T})^2} - \frac{2\left(\phi \bar{\phi} + \phi \phi f(U, \bar{U})\right)}{(T + \bar{T})^3},
\quad K_{T \bar{U}} = - \frac{\phi \phi \partial_{\bar{U}} f}{(T + \bar{T})^2}, \quad
K_{U \bar{U}} = \mc{K}_{U \bar{U}} + \frac{\phi \phi \partial_{\bar{U}} \partial_U f}{(T + \bar{T})}.
$$
Focusing on the $T$ and $U$ components, this gives
\be
\mc{K}^{-1} = \left( \begin{array}{cc} \frac{(T + \bar{T})^2}{3} + \mc{O}(T + \bar{T}) & - \frac{\phi \phi \partial_{\bar{U}} f}{3 \mc{K}_{U \bar{U}}}
\\ - \frac{\phi \phi \partial_{\bar{U}} f}{3 \mc{K}_{U \bar{U}}} & \mc{K}^{U \bar{U}} + \mc{O}(\frac{1}{T + \bar{T}}) \end{array} \right).
\ee
We again have $\eta \sim (T + \bar{T})^{-2}$, and the moduli spaces are approximately factorised at large volume: non-zero
$F^U$ cannot be induced from non-zero $F^T$. The restoration of factorisation at large volumes is consistent with intuition.
The breaking of factorisation occurred because the D3-brane positions mix with the K\"ahler moduli controlling 4-cycle volume.
This occurs because the D3 branes back-react on the geometry, and this backreaction alters the 4-cycle sizes \cite{hepth0607050}, which enters
into the holomorphic chiral superfields.
However, the larger the volume the more the effect of D3 brane back-reaction is diluted by the volume of the
compact space and the less effect it has on the moduli space factorisation.

Another source of corrections that violate factorisation are loop corrections. These can arise from either D3 or D7 branes.
We focus on the corrections due to D7 branes since these are more generic - it is always possible to avoid including D3 branes
by saturating the O3 tadpole with 3-form flux. Loop corrections to the K\"ahler potential in torodial backgrounds
have been computed in \cite{hepth0508043}. For convenience we fix $T_1 = T_2 = T_3$, $U_1 = U_2 = U_3$. The loop-corrected
K\"ahler potential due to the presence of D7 branes is then
\be
K = - \ln (S + \bar{S}) -3 \ln (T + \bar{T}) - 3 \ln (U - \bar{U}) + \frac{3}{256 \pi^6} \frac{\mc{E}_2(0, U)}{(T + \bar{T})^2}
\ee
Here $$
\mc{E}_2(0, U) = - \sum_{(m,n) \neq (0,0)} \frac{1920 (U - \bar{U})^2}{(n + mU)^2 (n+m \bar{U})^2}
$$
This Eisenstein series has the property that
$$
\partial_U \partial_{\bar{U}} E_2 (0, U) = - \frac{2}{(U - \bar{U})^2} \mc{E}_2(0,U).
$$
The K\"ahler metric is
\be
\mc{K} = \left( \begin{array}{cc} \frac{3}{(T + \bar{T})^2} & 0 \\ 0 & \frac{-3}{(U - \bar{U})^2} \end{array} \right)
+ 3 \left( \begin{array}{cc} \frac{6}{256 \pi^6} \frac{\mc{E}_2(0,U)}{(T + \bar{T})^4} & \frac{-2}{256 \pi^6}
\frac{\partial_{\bar{U}} \mc{E}_2(0, U)}{(T + \bar{T})^3} \\ \frac{-2}{256 \pi^6}
\frac{\partial_{\bar{U}} \mc{E}_2(0, U)}{(T + \bar{T})^3} & \frac{-2}{256 \pi^6}
\frac{ \mc{E}_2(0, U)}{(U - \bar{U})^2 (T + \bar{T})^2} \end{array} \right).
\ee
This gives
\bea
K^{T \bar{T}} & = & \frac{(T + \bar{T})^2}{3} + \ldots ,\\
K^{U \bar{T}} & = & \frac{1}{3} \frac{1}{128 \pi^6} \frac{1}{(T + \bar{T})} \left( \sum_{(n,m) \neq (0,0)}
\frac{2 (U - \bar{U})^3 \ti 1920}{(n + m U)^3 (n + m \bar{U})} \right).
\eea
The ratio
$$
\eta = \frac{K^{U \bar{T}}}{K^{T \bar{T}}} = \frac{1}{128 \pi^6} \frac{1}{(T + \bar{T})^3}
\sum_{(n,m) \neq (0,0)} \frac{2 (U - \bar{U})^3 \ti 1920}{(n + m U)^3 (n + m \bar{U})}.
$$
The breaking of factorisation due to the loop corrections goes as $(T + \bar{T})^{-3}$ at large cycle volumes.

A final source of factorisation-violating corrections to the K\"ahler potential are those arising from higher $\alpha'$ corrections.
Fluxes couple to complex structure moduli, and so
$\alpha'$-corrections involving the fluxes will therefore lead to corrections to the K\"ahler
potential that will mix K\"ahler and complex structure moduli, violating factorisation. For example, the 10d $\alpha'^3$ correction
$$
\frac{1}{\alpha'^4} \int d^{10} x \left( \mc{R} + \alpha'^3 \left( G_3^2 \mc{R}^3 + c.c \right) \right)
$$
should lead to a correction to $K$. The tree-level moduli kinetic terms arise from the dimensional reduction
of the $\mc{R}$ term. Using simple scaling arguments (flux is quantised on 3-cycles, so $G_3^2 \sim N^2/\mc{V}$,
while $\mc{R} \sim \mc{V}^{-1/3}$), it follows
that corrections due to the $G_3^2 \mc{R}^3$ term are suppressed by $\mc{V}^{-5/3}$ compared to the tree level kinetic terms.
At large volume, the violation of factorisation due to such terms will therefore be very small.

Let us summarise the results of this section. At leading order, the string theory moduli space factorises. The factorisation is
broken by loop corrections, $\alpha'$ corrections and by the presence of branes. In all these cases the breaking of factorisation is
suppressed at large volume: in the limit that the volume increases while all other fields are held constant factorisation is restored.

\subsection{Factorisation of Superpotential Yukawa Couplings}
\label{sec32}

The third requirement of mirror mediation is that the superpotential Yukawa couplings depend
only on the $\chi$ (flavour) sector while the
gauge couplings depend on the susy-breaking $\Psi$ sector.

An important feature of the definitions
of moduli superfields in string theory is the presence of Peccei-Quinn symmetries. In IIB these correspond
to $T \to T + i \epsilon$, $S \to S + i \epsilon$, with $U$ not having a shift symmetry, whereas in IIA all three sets of moduli
have shift symmetries, $S \to S + i \epsilon$, $T \to T + i \epsilon$, $U \to U + i \epsilon$.
These shift symmetries arise because the imaginary parts originate
from axionic terms: $\rm{Im}(T)_{IIB} = C_4, \rm{Im}(S)_{IIB} = C_0,
\rm{Im}(S)_{IIA} = C_3, \rm{Im}(T)_{IIA} = B_2, \rm{Im}(U)_{IIA} = C_3$. Axions only have topological couplings and so perturbation
theory (which is an expansion about topologically trivial states) is insensitive to them. The axionic shift symmetries can be broken
only by effects non-perturbative in the worldsheet ($\alpha'$) or spacetime ($g_s$) expansions.

In IIB, the $T$ shift symmetry is unbroken in both space-time and
world-sheet perturbation theory. It can be broken by D3-instantons.
The $S$ shift symmetry is unbroken in space-time perturbation theory and can be broken by D(-1)-instantons.
For IIA models, the $T$ shift symmetry is unbroken in world-sheet perturbation theory and is broken by worldsheet instantons,
whereas the $S$ and $U$ symmetries are unbroken in both spacetime and worldsheet perturbation theory and can only be broken by
D2-instantons.

Up to such non-perturbative effects the Peccei-Quinn symmetries remain exact.\footnote{Some axionic symmetries
can also be broken by fluxes; for example the IIB dilaton shift symmetry and the IIA K\"ahler moduli shift
symmetries can be broken by fluxes.
The breaking of shift symmetries
by fluxes, where it occurs, will not affect the arguments given here.}
This strongly constrains the moduli that can enter the superpotential and in particular the
superpotential Yukawa couplings.
The requirement that the superpotential be both holomorphic in the moduli and preserve the Peccei-Quinn symmetries
eliminates any perturbative dependence on moduli having the Peccei-Quinn symmetry.
Assuming the string coupling to be small, we then know that within spacetime
perturbation theory
\bea
\label{eq1}
Y_{\alpha \beta \gamma, IIA}(S, T, U) & = & Y_{\alpha \beta \gamma}(T), \\
\label{eq2}
Y_{\alpha \beta \gamma, IIB}(S, T, U) & = & Y_{\alpha \beta \gamma}(U).
\eea
The Yukawa couplings depend only on the T-moduli in IIA and on the U-moduli in IIB.
This structure matches onto the third assumption required for mirror mediation.
We can also now identify the $\Psi$ sector with the K\"ahler moduli in IIB and with the complex structure moduli in
IIA, and vice-versa for the $\chi$ sector.\footnote{Similar Peccei-Quinn symmetries constraining the
perturbative appearance of $T$ and $S$ moduli can be found in heterotic string theory.
The strongly coupled heterotic string is related by dualities to type I (that is, type IIB orientifold) models. This suggests that
the mirror mediation structure may also exist in heterotic models. Dilaton domination can be seen as an example of this,
where the susy breaking sector is viewed as the $(S,T)$ moduli and the $U$ moduli as the flavour sector.}

This structure of Yukawa couplings fits with the explicit computations (we will discuss these further below).
In IIB compactifications, Yukawa couplings have essentially classical origins. Chiral fermions arise from
the reduction of the DBI/super Yang-Mills actions in the presence of magnetic flux. The fermion modes and wavefunctions are found by
solving the higher-dimensional Dirac equation in the presence of magnetic flux.
The Dirac equation depends on the complex geometry - i.e. the complex structure - of the Calabi-Yau.
The Yukawa couplings arise from the classical overlap of these wavefunctions, and are non-vanishing even in the field theory
limit $g_s \to 0, \alpha' \to 0$. This is manifest from the form of (\ref{eq1}): the Yukawa couplings depend on the $U$ moduli,
which enter into neither the $g_s$ nor $\alpha'$ expansions.

In IIA, chirality arises through the pointlike intersection of D6-branes in extra dimensions.
Each intersection point gives rise to a chiral fermion.
Matter is localised
at the intersection point and so there are no classical couplings between different species.
Due to the spatial separation, all Yukawa couplings must arise
nonperturbatively. The Yukawa couplings are generated by worldsheet instantons, which
 are non-perturbative in the
$\alpha'$ expansion. They depend only on the K\"ahler moduli and appear as $e^{-2 \pi T_i}$, breaking the
$T$ Peccei-Quinn symmetry.

The superpotential
factorisation of equations (\ref{eq1}) and (\ref{eq2}) is broken by brane instantons, effects which are non-perturbative
in the string coupling. These allow a dependence on $T (U)$ moduli to appear in the IIB (IIA) superpotential or Yukawa couplings.
Such effects are non-perturbative in both the $g_s$ and $\alpha'$ expansions.
In the limit that either the string coupling is small or the volume is large such
effects are therefore exponentially suppressed. The fields entering the exponents are furthermore the same fields that
define the high-scale Standard Model gauge couplings. As these couplings are small
instanton effects are expected to be insignificant: in moduli stabilised models, it is often the case
that $e^{-T} \sim \frac{m_{3/2}}{M_P} \lesssim 10^{-15}$.

Moduli also have different roles in determining the gauge couplings.
Gauge couplings correspond to the volumes of cycles wrapped by branes.
For IIB with matter on D7 branes, which is the phenomenologically interesting case,
\be
f_a = T_a + h_a(F)S.
\ee
$h_a(F)$ depends on the magnetic fluxes present on the branes.
For IIA models with intersecting D6-branes,
\be
f_a = U_a,
\ee
where $U_a$ is the modulus (either complex structure or dilaton) controlling the volume of
the 3-cycle wrapped by the D6 branes.

It then follows that the third assumption of mirror mediation is satisfied in string compactifications, as the superpotential
Yukawa couplings depend only on one class of moduli whereas the gauge couplings depend on the other class.

\subsection{Factorisation of Physical Yukawa Couplings}

The fourth requirement of mirror mediation is that the matter metrics - and hence the physical Yukawa couplings -
factorise, with the $\chi$ sector determining the flavour structure and the $\Psi$ sector serving only as an overall normalisation.

To illustrate the different roles played by K\"ahler and complex structure moduli, we write down the full
expressions for the physical Yukawa couplings for toroidal compactifications.
In IIB compactifications
the classical Yukawa couplings - those applicable at leading order in the
 $g_s$ and $\alpha'$ expansions - can be computed in field theory. This procedure was carried out in the paper
 \cite{hepth0404229}, whose authors dimensionally reduced the ten-dimensional Yang-Mills action on a toroidal background in the presence
 of magnetic flux. This is related by T-duality to D3-D7 systems.
 The following expression was obtained for the physical Yukawa couplings:
\be
\label{fieldtheory}
Y_{ijk} = \underbrace{\frac{1}{\sqrt{\mc{V}}}}_{T-dependence} \underbrace{g_s \left( \prod_{r=1}^3 2 \rm{Im} U^r \right)^{\frac{1}{4}}
\left\vert \frac{\tilde{I}_1^r \tilde{I}_2^r}{\tilde{I}_1^r + \tilde{I}_2^2} \right\vert^{1/4}
\vartheta \left[ \begin{array}{c} \delta_{ijk}^{r} \\ 0 \end{array} \right] \left(0,
U^{r} \vert I_{ab}^r I_{bc}^r I_{ca}^r \vert \right)}_{U-dependence}
\ee
The quantities $I^r, \tilde{I}^r$ are integers related to flux quantisation, while
$\delta_{ijk}^r = \frac{i^r}{I^r_{ab}} + \frac{j^r}{I^r_{ca}} + \frac{k^r}{I^r_{bc}}$.
The flavour structure of the Yukawa couplings appear through the $\vartheta$-function. This depends on the $U$-moduli
whereas the K\"ahler moduli appear as an overall flavour-independent normalisation. The K\"ahler moduli can only affect the
overall scale of the Yukawa couplings and cannot affect the texture and relative hierarchies of the couplings.

For toroidal compactifications the Yukawas can also be computed in the full string theory.
The stringy computation for IIB with magnetised D9-branes gives \cite{hepth0404134, hepth0512067, hepth0610327, hepth0701292, 07091805}
\bea
\label{fullYukawas}
Y_{ijk}^{IIB} & = & \overbrace{\frac{1}{\mc{V}^{1/4}} \prod_{r=1}^3 \sigma_{abc}  \left( \frac{\Gamma(1 - \frac{1}{\pi} \phi^r_{ab})
\Gamma(1 - \frac{1}{\pi} \phi^r_{ca})\Gamma(\frac{1}{\pi} (\phi^r_{ab} + \phi^r_{ca}))}{(2 \pi)^3 \Gamma(\frac{1}{\pi} \phi^r_{ab})
\Gamma(\frac{1}{\pi} \phi^r_{ca})\Gamma(1 - \frac{1}{\pi} (\phi^r_{ab} + \phi^r_{ca}))} \right)^{1/4}}^{\textrm{T-dependence}} \nonumber \\
& & \ti e^{\phi_{10} /2} \underbrace{(U^r)^{1/4} \vartheta \left[ \begin{array}{c}
\delta_{ijk}^r \\ 0 \end{array} \right] \left(0; U^r I^r_{ab} I^r_{bc} I^r_{ca} \right)}_{\textrm{U-dependence}}.
\eea
The angles $\phi^r_{ab}$ satisfy $\phi^r_{ab} + \phi^r_{bc} + \phi^r_{ca} = 0$ and are given by
\be
\phi^r_{ab} = \arctan \left( \frac{f_b^r}{t_2^r} \right) - \arctan \left( \frac{f^r_a}{t_2^r} \right),
\ee
with $t_2$ 2-cycle volumes.
In the small-angle dilute flux limit, $t_2 \gg f_a, f_b$, this expands as
\be
\phi^r_{ab} = \left( \frac{f_b^r - f_a^r}{t_2^r} \right) - \frac{1}{3} \left( \left( \frac{f_b^r}{t_2^r} \right)^3
- \left( \frac{f_a^r}{t_2^r} \right)^3 \right) + \frac{1}{5} \left( \left( \frac{f_b^r}{t_2^r} \right)^5
- \left( \frac{f_a^r}{t_2^r} \right)^5 \right) + \ldots
\ee
The expansion of the gamma functions gives
\be
\label{gammexp}
\left( \frac{\Gamma(1 - \frac{1}{\pi} \phi^r_{ab})
\Gamma(1 - \frac{1}{\pi} \phi^r_{ca})\Gamma(\frac{1}{\pi} (\phi^r_{ab} + \phi^r_{ca}))}{(2 \pi)^3 \Gamma(\frac{1}{\pi} \phi^r_{ab})
\Gamma(\frac{1}{\pi} \phi^r_{ca})\Gamma(1 - \frac{1}{\pi} (\phi^r_{ab} + \phi^r_{ca}))} \right)^{1/4}
= \frac{1}{\pi}
\frac{\phi^r_{ab} \phi^r_{ca} }{\phi^r_{ab} + \phi^r_{ca}} - \frac{2}{\pi^4} \left( \phi^r_{ac} \phi^r_{ca} \right)^2 + \ldots...
\ee
In the limit that the angles $\phi_{ab} \to 0$, equation (\ref{fullYukawas}) reduces to (\ref{fieldtheory}).
Note however that eq. (\ref{fullYukawas}) retains the factorised form to all orders in $\alpha'$.
This also shows that the breakdown of the single classical $T$-scaling occurs at
order $\left( \frac{f}{t} \right)^2$ in the dilute flux expansion - this is an $\mc{O}(\alpha'^2)$
effect.\footnote{This also implies that in the field theory limit the supersymmetry condition
$\phi_1 + \phi_2 + \phi_3 = 0$ does not place constraints on the K\"ahler moduli.}
The breakdown of the classical scaling can be understood from the fact that the higher dimensional action is actually the DBI action
rather than the super Yang-Mills.

In the T-dual picture with intersecting D6-branes, the angles $\phi^r_{ab}$ are the physical intersection angles
between different branes.
The IIA result for intersecting D6-branes is the mirror-symmetric form of the above in which $T$ and $U$ are interchanged.
It takes the form \cite{hepth0303083}
\bea
\label{fullYukawasIIA}
Y_{ijk}^{IIA} & = & e^{\Phi_4 /2} \prod_{r=1}^3 \sigma_{abc} (t^r)^{1/4} \left( \frac{\Gamma(1 - \frac{1}{\pi} \phi^r_{ab})
\Gamma(1 - \frac{1}{\pi} \phi^r_{ca})\Gamma(\frac{1}{\pi} (\phi^r_{ab} + \phi^r_{ca}))}{(2 \pi)^3 \Gamma(\frac{1}{\pi} \phi^r_{ab})
\Gamma(\frac{1}{\pi} \phi^r_{ca})\Gamma(1 - \frac{1}{\pi} (\phi^r_{ab} + \phi^r_{ca}))} \right)^{1/4} \nonumber \\
& & \ti \vartheta \left[ \begin{array}{c}
\delta_{ijk}^r \\ 0 \end{array} \right] \left(0; t^r I^r_{ab} I^r_{bc} I^r_{ca} \right).
\eea
Here $\Phi_4 = \phi_{10} - \half \ln \mc{V}$ is the 4-dimensional dilaton and
the $t^r$ of IIA refer to 2-cycle volumes. In IIA string theory, the angles $\phi_{ab}$ depend on the complex structure moduli rather than
the K\"ahler moduli as in IIB.

From the above formulae we see that the Yukawa couplings do admit a factorised form.
The flavour-dependent part is encoded in the $\vartheta$-functions and depends on the complex structure in IIB and
on the K\"ahler moduli in IIA. These involve exponentials, which could naturally lead to the mass hierarchies within the
Standard Model.

In addition, there is a universal flavour-independent normalisation prefactor. This depends on
the K\"ahler moduli in IIB and on the complex structure moduli in IIA. In the dilute flux limit in which $\phi_{ab} \to 0$ for all
$\phi$, the K\"ahler moduli appear simply as an overall prefactor with a single power.
This is illustrated in the field theory limit through the formula (\ref{fieldtheory}).

The factorisation of the Yukawa couplings has a simple field theory origin
which is most easily understood in the IIB formalism. This also allows an understanding of why
factorisation extends beyond the toroidal case.
In IIB compactifications, chirality arises
due to the presence of magnetic fluxes on brane world-volumes. Chiral fermions arise as zero modes of the Dirac equation
in the presence of magnetic flux. The relative magnetic flux distinguishes two brane stacks and leads to bifundamental fermions.
Different flavours correspond to different zero modes. The Dirac equation is
\be
\label{Dirac}
\Gamma^M D_M \lambda = \Gamma^M \left( \partial_M \lambda + \frac{1}{4} \omega_M^{kl} \Sigma_{kl} \lambda
+ [A_M, \lambda] \right) = 0.
\ee
 The spin connection $\omega_M^{kl}$ is defined through the vielbein
$e_a^M$, with $g^{MN} = e_a^M \eta^{ab} e_b^N$, $\Gamma^M = e_a^M \gamma^a$ with $\gamma^s$ the flat-space Dirac matrices, and
$\Sigma_{kl} = \frac{1}{4} \gamma_{[k,}\gamma_{l]}$. Then
\be
\omega_M^{ab} = \half g^{RP} e_R^{[a} \partial_{[M} e_{P]}^{b]} + \frac{1}{4} g^{RP} g^{ST} e^{[a}_R e_T^{b]}
\partial_{[S} e_
{P]}^c e^d_M \eta_{cd}.
\ee
Under rescalings $g \to \lambda g$ the spin connection is unchanged. Likewise, the gauge field $A_i$ is specified in terms of the
complex coordinates of the space and is unaffected by rescalings of the metric.
Any zero mode of (\ref{Dirac}) is therefore unaltered by a metric rescaling.
As the texture of Yukawa couplings comes from the overlap of zero modes,
these are also unaffected by metric rescalings.

Rescalings do however affect the normalisation of the zero modes. To have canonical kinetic terms, the zero modes
must satisfy
\be
\label{cankin}
\int_{\Sigma} \sqrt{g} \vert \psi \vert^2 = 1.
\ee
where $\Sigma$ is the submanifold on which they are supported.
The wavefunctions must then be normalised as $\psi_N \sim
\frac{1}{\sqrt{Vol(\Sigma)}} \psi_0$. This normalisation depends only on the volume of $\Sigma$ and
is therefore flavour-independent.

Physical Yukawa couplings arise from the triple overlap of three
normalised wavefunctions (due to supersymmetry the bosonic zero modes
 have the same functional form as their fermionic partners). Metric rescalings enter the
 physical Yukawas through the normalisation condition (\ref{cankin}), which is flavour-independent.
 This gives a universal (i.e. factorised) form with respect to the metric modes which rescale cycle volumes:
 these modes are the K\"ahler moduli.

 Specifically, in the case that all chiral matter arises from branes wrapping the same cycle, the Yukawa couplings descend
 from the term
 $$
 \int_{\Sigma} \sqrt{g} \Gamma^\mu D_\mu \psi \to \int_{\Sigma} \sqrt{g} \bar{\psi} (\Gamma^M A_M) \psi.
 $$
 This applies both to D9 branes and to models of branes at (resolved) singularities.
 Canonical normalisation of the kinetic terms requires (up to numerical factors)
 $$
 \int_{\Sigma} \sqrt{g} (\bar{\psi} \psi)^2 = 1, \int_{\Sigma} \sqrt{g} (\Gamma^M A_M)^2 = 1,
 $$
and so putting all factors of the cycle volumes together the physical Yukawa couplings have a universal scaling of
$Vol(\Sigma) \ti \left( \frac{1}{\sqrt{Vol(\Sigma)}} \right)^3 = \frac{1}{\sqrt{Vol(\Sigma)}}$. This scaling
is independent of the detailed flavour structure.
This is precisely the behaviour seen in equation
(\ref{fieldtheory}), where all chiral matter arises from wrapped D9-branes.
However this argument does not rely on a toroidal background and applies equally well to
the Calabi-Yau case.

\subsection{Factorisation of Matter Metrics}

In the expression for the soft masses and A-terms it is not just the Yukawa couplings but actually the matter metrics that
enter the physical Yukawa couplings. These are not so easy to compute directly through dimensional reduction.
However, in the field theory limit the modular weights of the matter metrics
can be inferred indirectly from the scalings of the physical Yukawa couplings.
The supergravity structure implies the physical Yukawa couplings can be written as
\be
\hat{Y}_{\alpha \beta \gamma} = e^{\hat{K}/2} \frac{Y_{\alpha \beta \gamma}}{(\tilde{K}_{\alpha} \tilde{K}_{\beta}
\tilde{K}_{\gamma})^{\half}}.
\ee
The K\"ahler potential $\hat{K}$ is known. We have seen in section
\ref{sec32} that the combination of holomorphy and shift symmetries implies
certain moduli cannot appear in $Y_{\alpha \beta \gamma}$,
and so if the modular scaling of the physical Yukawa coupling can be computed then the
modular weights of the corresponding kinetic terms can be inferred.
This logic was used in \cite{hepth0609180}
to compute the modular weights for localised chiral D7-D7 matter in the large volume models.
For example, in the case above where all chiral matter arises from branes wrapping the same cycle, we can deduce that
\be
e^{\hat{K}/2} \frac{1}{(\tilde{K}_\alpha \tilde{K}_\beta \tilde{K}_\gamma)^{\half}} \sim \frac{1}{\hbox{Vol}(\Sigma)^{\half}}.
\ee
The derivation of this only uses the region on which the wavefunction (i.e. the cycle $\Sigma$)
is supported and not the particular form of the wavefunction.
Each index ($\alpha$, $\beta$ and $\gamma$) of the Yukawa couplings corresponds
to a different set of gauge-charged fields.
If a different flavour is used in computing the Yukawa couplings,
then this corresponds to replacing the index $\alpha$ by $\alpha'$ (e.g. using the top quark rather than the up quark).
As fields with identical gauge charges are supported on the same cycle
the same volume scaling occurs, which implies the two flavours must, under metric rescalings,
 have the same modular weights in the field theory limit.

The Peccei-Quinn symmetries $T \to T + i \epsilon$ also enforce the reality conditions that are present in assumptions 2 and 4 of mirror mediation.
The requirement that the shift symmetry be unbroken in perturbation theory implies that there can be no explicit dependence
on the (imaginary) axionic components of $T$: $T$ can only appear in the K\"ahler metrics as $(T + \bar{T})$.

The matter metrics have been explicitly computed in toroidal backgrounds using string worldsheet analyses. The
intersection of any two D6 branes is invariantly characterised by three angles $\theta_1$, $\theta_2$, $\theta_3$, with
$\theta_1 + \theta_2 + \theta_3 = 0$. In \cite{07052366} the expression for the matter metric for the chiral field
on the brane intersection is given by
\be
\label{matt1a}
K_{ij}^{ab} = \delta_{ij} S^{-1/4} \prod_{I=1}^3 U_I^{-1/4} T_I^{-(\half \pm \half \rm{sign}(I_{ab}) \theta^I_{ab})}
\left( \frac{\Gamma(\theta^1_{ab})\Gamma(\theta^2_{ab})\Gamma(1+ \theta^3_{ab})}{\Gamma(1 - \theta^1_{ab})
\Gamma(1 - \theta^2_{ab})\Gamma(1 - \theta^3_{ab})}\right)^{\half}
\ee
Here $S$, $T$ and $U$ are short for $S + \bar{S}$, $T + \bar{T}$ and $U + \bar{U}$.
Recall $\theta$ here depends solely on the $U$ moduli.
Converted to IIB, this expression gives
\be
\label{matt2a}
K_{ij}^{ab} = \delta_{ij} S^{-1/4} (T^1 T^2 T^3)^{-1/4} \prod_{I=1}^3 U_I^{-(\half \pm \half \rm{sign}(I_{ab}) \theta^I_{ab})}
\left( \frac{\Gamma(\theta^1_{ab})\Gamma(\theta^2_{ab})\Gamma(1- \theta^1_{ab} - \theta^2_{ab})}{\Gamma(1 - \theta^1_{ab})
\Gamma(1 - \theta^2_{ab})\Gamma(\theta^1_{ab} + \theta^2_{ab})}\right)^{\half}
\ee
The angles $\theta$ here depends only on the $T$ moduli and the product of gamma functions has the same expansion as in eq.
(\ref{gammexp}). This breaks the single modular weight at order $(f/t)^2$, namely at $\mc{O}(\alpha'^2)$.
In \cite{07090245}, the same expression is obtained, however without the $U_I^{-\half \rm{sign}(I_{ab}) \theta^I_{ab}}$ term.
If the term $U_I^{-\half \rm{sign}(I_{ab}) \theta^I_{ab}}$ is present, then expanding the power
it breaks the factorised form of the matter metric at
$\mc{O}(\alpha')$, that is $\mc{O}(f/t)$.
As $\theta_1 + \theta_2 + \theta_3 = 0$, this angular dependence of the $U$ moduli is not present in the form of the physical Yukawa couplings.

However, the factorised structure of the matter metrics that is the fourth assumption of mirror mediation is present at leading order.
The presence of a universal modular weight is certainly broken at $\mc{O}(\alpha'^2)$ by the
gamma function product. Depending on the existence
of the $U_I^{-\half \rm{sign}(I_{ab}) \theta^I_{ab}}$, this may also be broken at $\mc{O}(\alpha')$.

There is a further point of interest. For toroidal examples of IBWs, the matter metrics are actually entirely diagonal and
flavour-universal. This is easy to understand. The diagonal nature comes becuase different chiral fermions are located at
different intersection points, and so are physically separated in space. The metrics for these fields can
only be non-diagonal at a non-perturbative level through brane or worldsheet instantons. The flavour-universality occurs because in a torus the
intersection angles are universal between different flavours: the torus is flat and two hyperplanes always intersect at the
same angles. As it is the angles that determine the matter metric, it follows that on a torus the flavour-universality of the matter metrics
occurs to all orders in $\alpha'$ (note that this does not hold for species of different gauges charges - here the brane stacks intersect at different angles, giving different matter metrics).
We will see in appendix \ref{appendixsec} that for Calabi-Yau intersecting brane worlds the intersection angles can be family non-universal.
In this case the factorisation is expected to break at subleading order in the $\alpha'$ expansion.

\section{Supersymmetry Breaking}
\label{sec4}

Section \ref{sec3} has shown that the factorisation of the moduli space that is necessary to realise mirror mediation does
indeed occur in string compactifications. To realise mirror mediation, it is necessary that susy breaking
also respects this factorisation, with the goldstino lying dominantly in the sector mirror to that in which the flavour structure
originates. While it is possible to study the moduli space in both the IIA and IIB contexts,
the process of moduli stabilisation, supersymmetry breaking and hierarchy generation is much better understood for IIB
compactifications. In this section we
will therefore focus on type IIB models, as it is
possible to be explicit about the origin of supersymmetry spontaneously broken at hierarchically low energy scales.
We shall comment briefly on the IIA case at the end.

\subsection{The GKP Limit}

In type IIB models the flavour structure arises through the complex structure moduli.
If mirror mediation is to occur, supersymmetry must be dominantly broken in the K\"ahler moduli sector.
Fortunately this is naturally realised in IIB flux compactifications. Here
I shall only state results - detailed reviews of flux compactifications are
in \cite{hepth0509003, hepth0610102, hepth0610327, hepth0701050}.
We start with the models of Giddings, Kachru and Polchinski \cite{hepth0105097}, which
correspond to flux compactifications
of IIB string theory with D3 and D7 branes on Calabi-Yau orientifolds.
At leading order in $g_s$ and $\alpha'$, the low energy supergravity theory is described by
\bea
W & = & \int G_3 \wedge \Omega, \\
K & = & -2 \ln \left( \mc{V}(T_i + \bar{T}_i) \right) - \ln \left( i \int \Omega \wedge \bar{\Omega} \right) - \ln (S + \bar{S}).
\eea
$T_i$ are the K\"ahler moduli, and $\Omega$ depends on the complex structure moduli $U$. For now we neglect the possible
presence of additional brane moduli, but we shall remedy this below.
As is well known, this theory has no-scale structure and the scalar potential reduces to
\bea
V & = & \sum_{I,J = S,T,U} e^K \left( K^{I \bar{J}} D_I W D_{\bar{J}} \bar{W} - 3 \vert W \vert^2 \right) \nonumber \\
& = & \sum_{I,J=U,S} e^K K^{I \bar{J}} D_I W D_{\bar{J}} \bar{W} = \sum_{I,J=U,S} K_{I \bar{J}} F^I \bar{F}^{\bar{J}}.
\eea
This stabilises the dilaton and complex structure moduli in a supersymmetric fashion,
$$
D_{U_i} W = D_S W = F^{U_i} = F^S = 0, \qquad \forall i.
$$
The K\"ahler moduli are unstabilised and break supersymmetry, $F^{T_j} \neq 0$. This realises the final requirement of
mirror mediation, as the supersymmetry breaking occurs in one of the mirror sectors, while flavour physics
is generated in the other sector.

In the GKP limit, the supersymmetry breaking actually has further structure and can be
 associated to the breathing mode.
This can be seen by identifying the goldstino.
The goldstino is the fermionic mode eaten by the massive gravitino as supersymmetry is broken.
The goldstino parametrises the direction of
supersymmetry breaking, and it is the couplings of this direction to matter that determine the
supersymmetric soft terms.

In general the Calabi-Yau volume is a function of many moduli,
$\mc{V} = \mc{V}(T_i + \bar{T_i}), i = 1, \ldots h^{1,1}$. As the $T_i$ parametrise
the volume of 4-cycles, the volume is
a homogeneous function of the $T_i$ of degree 3/2. To identify the goldstino,
at any fixed point $P$ in moduli space we can go to local
coordinates. We identify and label the goldstino through the bosonic mode that the goldstino is the fermionic
counterpart to. As the K\"ahler potential is a function only of $T_i + \bar{T}_i$ and not of the axionic components
$c_i = \hbox{Im}(T_i)$, it is sufficient to focus on the real parts
$\tau_i = \hbox{Re}(T_i)$ of the fields. $W$ is independent of the $T$-moduli, implying
$$
D_{T_j} W = \partial_{T_j} W + (\partial_{T_j} K) W = (\partial_{T_j} K) W = \half (\partial_{\tau_j} K) W.
$$
 For any fields $\psi, \phi \in \{ \tau_i \}$,
\bea
\partial_{\psi} K & = & - 2 \frac{\partial_\psi \mc{V}}{\mc{V}}, \\
\partial_{\psi} \partial_{\phi} K & = & -2 \frac{\partial_{\psi} \partial_{\phi} \mc{V}}{\mc{V}} + 2\frac{\partial_\psi \mc{V}
\partial_\phi \mc{V}}{\mc{V}^2}.
\eea
At any point in moduli space $P \in {\tau_i}$ we go to local coordinates $\{\tau_i\} = \{\tau_{i,0}\} + (X, Y_i)$,
in which the $Y$-directions are transverse
to the overall volume, $\partial_{Y_i} \mc{V} \vert_P = 0$, implying that $D_{Y_i} W = 0$.
The $X$ direction is defined to be the overall scaling direction
$\tau_i \to (1 + X) \tau_i$, and so $D_X W \neq 0$.
Evaluating the K\"ahler metric we get
$$
\Big( \partial_{Y_i} \partial_{X} V \Big) \Big\vert_P  =  -2 \Big( \frac{\partial_{Y_i} \partial_{X} \mc{V}}{\mc{V}} \Big)
\Big\vert_{P} + 2 \Big(\frac{\partial_{Y_i} \mc{V}
\partial_{X} \mc{V}}{\mc{V}^2} \Big) \Big\vert_P.
$$
As $X$ is precisely the overall scaling mode $\tau_i \to (1 + X) \tau_i$ and $\mc{V}$ is homogeneous in the $\tau_i$
of degree 3/2, $\partial_X \mc{V} =  \frac{3 \mc{V}}{2}$. Therefore
$$
\partial_{Y_i} \partial_{X} V \vert_P = \frac{3}{2} \partial_{Y_i} \mc{V} \vert_P = 0,
$$
as the $Y_i$ directions are defined to be transverse to the overall volume at $P$.
It follows that  $K_{X Y_i}\Big\vert_P = 0$. This implies the off-diagonal elements of the K\"ahler metric vanish:
$$
K = \left( \begin{array}{c|c} K_{X\bar{X}} & 0 \\ \hline 0 & K_{Y_i \bar{Y}_j} \end{array} \right).
$$
As $D_{Y_i}W =0$ and $D_{X} W \neq 0$, it follows that $F^X \neq 0$ and $F^{Y_i} = 0$.
The $X$-direction thus corresponds to the goldstino - it is both orthogonal to all other directions in moduli space
and the only direction to break supersymmetry. The $X$ direction corresponded to the overall
rescaling mode $\tau_i \to \lambda \tau_i$. In terms of the Calabi-Yau metric, this is simply the
breathing mode
$$
g_{i \bar{j}} \to \lambda^2 g_{i \bar{j}},
$$
appropriately complexified with axionic fields.

The identification of the Goldstino with the breathing mode extends to the
case where D3 and D7 position modes are included and there are brane moduli that mix with the
K\"ahler and complex structure moduli. In this case the correct expression for the holomorphic chiral superfields becomes rather complicated.
(cf eq. (\ref{d3red}) or  eq. (4.35) of \cite{hepth0409098}).
The K\"ahler moduli no longer simply involve the complexified cycle sizes of the Calabi-Yau, but also
mix with (for example) D3 position moduli, D7 position moduli and Wilson line
moduli. Nonetheless the K\"ahler potential is still given by
$$
K = - 2 \ln \left( \mc{V}(T + \bar{T}, \phi_{D3}, \phi_{D7}, A_{WL}, \ldots) \right) -
\ln \left( \int \Omega \wedge \bar{\Omega} \right) - \ln (S + \bar{S}),
$$
where $\mc{V}$ is the physical volume of the Calabi-Yau. In terms of the physical cycle sizes $\tau_i$, the volume $\mc{V}$
remains a homogeneous polynomial of degree 3/2. However it is no longer possible to
write $\tau_i = \hbox{Re}(T_i)$.

The argument above giving the goldstino direction nonetheless still applies.
At any point in moduli space, we can go into local coordinates $(X, Y_i)$, where the
$Y$-directions are transverse to the overall volume, with $\partial_{Y_i} \mc{V} = 0$ and thus $D_{Y_i} W = 0$.
The $Y_i$ directions now include brane motions as well as directions in K\"ahler moduli space.
The $X$-direction again corresponds to the overall rescaling
$\tau_i \to (1 + X) \tau_i$. As $\partial_X \mc{V} \propto \mc{V}$, we again have $K_{X Y_i} = 0$. This implies the $X$ direction is
transverse to all $Y$ directions and thus the goldstino is the breathing mode
$$
\tau_i \to (1 + X) \tau_i, \qquad g_{i \bar{j}} \to \mu^2 g_{i \bar{j}}.
$$
This result is not so surprising, given the expression for the gravitino mass.
For vanishing cosmological constant the gravitino mass is
the order parameter of supersymmetry breaking. It is given by
\be
\label{gravitinomass}
m_{3/2} = e^{K/2} W = \frac{W}{\mc{V}}.
\ee
As $W$ depends only on the complex structure moduli which have been fixed,
we see that the goldstino should correspond to a rescaling of the overall volume.

The fact that the goldstino aligns with the breathing mode makes it manifest that
 the resulting soft masses will be flavour-universal up to possible corrections of higher order in the $\alpha'$
 and $g_s$ expansions. The resulting soft masses are determined by the coupling of the breathing mode
 to the different flavours.
From the discussion in section
\ref{sec3}, in the field theory limit the size moduli only enter the
 Yukawa couplings through a flavour-blind normalisation factor,
 characterised by the modular weight.  Metric rescalings affect the normalisation, but not the structure, of the Yukawa couplings.
 The coupling of the goldstino to the chiral matter is therefore
 only sensitive to the modular weight and not to the detailed
 flavour structure.

The GKP models therefore give a very attractive pattern of supersymmetry breaking, which ensures that
soft masses will be flavour-universal. However, as these models only stabilise the dilaton and complex structure moduli
they are incomplete.
We would like to retain the supersymmetry breaking structure of the GKP models
while also stabilising the K\"ahler moduli.

In realistic phenomenological models of supersymmetry breaking, it is furthermore
necessary that the gravitino be hierarchically small with a mass
 close to the TeV scale. From (\ref{gravitinomass}), it is clear that this can be accomplished in one of two ways:
either the superpotential $W$ is extremely small, $W \sim 10^{-15}$, or the volume $\mc{V}$ is extremely large
$\mc{V} \sim 10^{15}$. We now consider moduli-stabilised cases where
these properties are realised.
These cases correspond respectively to the KKLT and large volume frameworks for moduli-stabilisation.

\subsection{Models with full moduli stabilisation}

We consider two proposals for low energy supersymmetry together with moduli stabilisation
in IIB flux models.

\subsubsection{KKLT}

In KKLT \cite{hepth0301240}, non-perturbative effects
(from either Euclidean D3-instantons or gaugino condensation)
are introduced in each K\"ahler modulus. In the low energy theory these induce a
superpotential
\be
W = W_0 + \sum_i A_i e^{-a_i T_i}.
\ee
A hierarchically small gravitino mass, required to stabilise the gauge hierarchy,
can be obtained by tuning the tree-level flux superpotential $W_0$
to very small values. At the current level of understanding there is no dynamical explanation for why the
flux superpotential should be small rather than taking $\mc{O}(1)$ values.
The K\"ahler moduli are stabilised supersymmetrically by solving the equations
$D_{T_i} W = 0$. In a supersymmetric vacuum all the soft terms vanish and the vacuum
energy is AdS, with $F^U = F^T = 0$ and
$$
V_0 = -3 m_{3/2}^2 M_P^2.
$$
As the AdS vacuum is supersymmetric,
this step destroys the factorised structure of supersymmetry breaking present in GKP.
To match the cosmological constant it is necessary to include a source of susy-breaking energy to uplift
the AdS vacuum to Minkowski.
The resulting phenomenology of
supersymmetry breaking is determined entirely by the origin and nature of this uplifting, and different
methods can give quite different results.

Generally, we imagine uplifting with an additional hidden sector. The hidden sector can arise
from anti-D3 branes, fluxes or strong gauge dynamics.
At the level of the scalar potential for the K\"ahler moduli, the precise origin is not so important and
the uplifting can be parametrised by adding a term
$$
V = \sum_{i=T} e^K \left( K^{i \bar{j}} D_i W D_{\bar{j}} \bar{W} - 3 \vert W \vert^2 \right) + \frac{\epsilon}{\mc{V}^{2}}.
$$
The dependence on the volume comes from the universal $e^K$ term in the scalar potential.
In certain cases $\epsilon$ may also depend on the volume; for example when uplifting is performed using a
warped throat then $\epsilon$ has an effective dependence of $\mc{V}^{2/3}$, giving an overall dependence of $\mc{V}^{-4/3}$.
So long as such an uplift term depends solely on the volume,
directions $Y$ transverse to the volume will still be extremised
at $D_Y W = 0$, and restricting to the K\"ahler moduli alone, the
dominant F-term will still align with the overall breathing mode.
However it is easy to check \cite{hepth0411066} that the
magnitude of the resulting F-term for the $T$ fields is
$
F^T \sim \frac{m_{3/2}}{\ln (m_{3/2})}.
$
As
$$
(F^T)^2 \sim \frac{m_{3/2}^2}{\ln(M_P/m_{3/2})^2} \ll 3 m_{3/2}^2 M_P^2,
$$
this implies the $T$-fields only ever play a subdominant role in
susy breaking, in contrast to the GKP case. It is therefore
not possible for KKLT models to realise mirror mediation in its pure sense,
as the $T$ fields can never dominate supersymmetry breaking.

The soft terms in a KKLT framework are determined by the details and couplings of the uplifting sector.
For example, suppose the uplifting is carried out using a metastable susy breaking vacuum in
the complex structure moduli sector as
proposed in \cite{hepth0402135}. In this case the dominant F-terms are located in the complex structure moduli
sector, giving
$$
F^U \sim  m_{3/2}, \qquad F^T \sim \frac{m_{3/2}}{\ln m_{3/2}}.
$$
This is bad news from the viewpoint of flavour physics, as the dominant F-term lies in precisely the same sector from
which flavour physics originates. This is expected to
lead to large unobserved FCNCs and CP violation,
generated for example through the $F^U \partial_U Y_{\alpha \beta \gamma}(U)$ contribution to the A-terms.
The flavour constraints may be mitigated through the large scalar masses, but this is at tension with the
requirement of naturalness in the Higgs sector.
Similar expressions apply for uplifting with a matter sector.

If the matter metrics of the Standard Model fields do not depend on
the hidden sector fields, then the
$F^{hidden}$ terms do not contribute to scalar masses. In this case
the scalar masses come solely from the universal
$m_{3/2}^2$ supergravity term in (\ref{scalarmass}).
In this case the soft scalar masses are heavier than the gauginos by a factor
$\ln(M_P/m_{3/2})$,
$$
m_i^2 \simeq m_{3/2}^2, \qquad M_i \sim \frac{m_{3/2}}{\ln (M_P/m_{3/2})}.
$$
The soft masses are now so heavy that flavour constraints are less significant. However, given the
direct search limits on gaugino masses, the heaviness of the scalar
masses is in tension with naturalness in the Higgs sector.

In order to satisfy the flavour constraints with scalar and gaugino masses of comparable order,
it is necessary that the supersymmetry breaking that cancels the vacuum energy
be both insensitive to CP and flavour and also cancel the
universal supergravity $m_{3/2}^2$ contribution to the scalar masses.
This is equivalent to the statement that the uplifting supersymmetry breaking is sequestered from the visible sector.
If this occurs, both scalar and gaugino masses are dominantly determined by the $F^T$ components, together with
anomaly-mediated contributions that are of size
$$
m_{anomaly} \sim \frac{m_{3/2}}{8 \pi^2}  \sim \frac{m_{3/2}}{\ln (M_P/m_{3/2})}.
$$ This is the mirage mediation
scenario \cite{hepth0503216, hepph0504036, hepph0504037}.

\subsubsection{LARGE volume models}

KKLT stabilisation only incorporates non-perturbative superpotential corrections to the GKP framework.
The additional inclusion of perturbative $\alpha'$ corrections to the
K\"ahler potential gives rise to the large volume models
\cite{hepth0502058, hepth0505076}.
Somewhat unexpectedly, the inclusion of $\alpha'$ corrections to the scalar potential generates a new
supersymmetry-breaking minimum at exponentially large volumes. The reason why $\alpha'$ corrections can affect
the potential at such large volumes is that the tree-level potential vanishes due to the
no-scale structure,
making the $\alpha'$ corrections the leading perturbative
contributions to the scalar potential. These minima exist for all values of the
flux superpotential $W_0$. In addition to improved technical control,
the appearance of exponentially large volumes gives
a dynamical generation of the weak hierarchy.

The simplest example of these models is for the $\mbb{P}^4_{[1,1,1,6,9]}$ Calabi-Yau,
the volume of which is $\mc{V} = \tau_b^{3/2} - \tau_s^{3/2}$. The supergravity theory is
\bea
K & = & - 2 \ln \left( \left(\frac{T_b + \bar{T}_b}{2} \right)^{3/2} - \left(\frac{T_s + \bar{T}_s}{2} \right)^{3/2} + \frac{\xi}{g_s^{3/2}} \right), \\
W & = & W_0 + A_s e^{-a_s T_s}.
\eea
$\xi$ parametrises the $\alpha'$ correction. This theory can be shown to have
a supersymmetry breaking minimum at $\mc{V} \sim W_0 e^{a_s \tau_s}$, with
$\tau_s \sim \xi^{2/3}/g_s$. There are two cycles, one large ($\tau_b$) and one small ($\tau_s$).
The large cycle controls the overall volume and the small cycle the size of a blow-up mode. The standard model is localised
on D7 branes wrapping the blow-up mode.
The phenomenology of this model with regard to low-energy supersymmetry has been studied in \cite{hepth0505076, hepph0512081, hepth0605141,
hepth0610129, 07043403, 07040737}.

Before uplifting, the AdS vacuum energy is
$$
V_{AdS} \sim m_{3/2}^3 M_P \ll m_{3/2}^2 M_P^2.
$$
The fact that $V_{AdS}$ is hierarchically smaller than the natural supergravity scale
 of $m_{3/2}^2 M_P^2$ indicates
that the no-scale structure present in GKP survives to leading order:
the vacuum energy is much less than the scale of
susy breaking would suggest. This can also be seen in the mass of the volume modulus, which is \cite{hepth0505076}
$$
m_{Vol} \sim m_{3/2} \left( \frac{m_{3/2}}{M_P} \right)^{\half} \ll m_{3/2}.
$$
The lightness of the volume modulus and the smallness of the vacuum energy are both residues of the tree-level no-scale
structure. The volume modulus is the pseudo-Goldstone boson of the tree-level scaling symmetry $g \to \lambda g$,
which is broken by $\alpha'$ corrections.
The F-terms are \cite{hepth0505076, hepth0610129}
$$
F^b \sim m_{3/2} M_P, \qquad F^s \sim m_{3/2}^{3/2} M_P^{\half}.
$$
The goldstino aligns
dominantly with the overall breathing mode and
the susy breaking of the large volume models is largely inherited from that of the GKP case.

Additional sources of supersymmetry breaking are necessary to cancel the negative vacuum energy.
The required magnitude of this is $(F^{uplift})^2 \sim m_{3/2}^3 M_P \ll
(F^T)^2 \sim m_{3/2}^2 M_P^2$, and so the magnitude of soft terms due to the uplifting energy is
$$
m_{soft} \sim \frac{F^{uplift}}{M_P} \sim m_{3/2} \left( \frac{m_{3/2}}{M_P} \right)^{\half}.
$$
For $m_{3/2} \sim 1 \hbox{TeV}$, $m_{soft} \sim 1 \hbox{MeV}$ and so this contribution is negligible.

Independent of the details of the uplift, the
goldstino lies dominantly in the K\"ahler moduli sector.
The large volume models thus realise mirror mediation, as the supersymmetry breaking
is always dominated by the K\"ahler moduli.
The goldstino is, up to a small correction of order $\mc{V}^{-1/2}$, still aligned with the overall breathing mode.
Matter is localised on D7 branes wrapping the blow-up cycle, which can be thought of as a hole in the bulk of the Calabi-Yau.
The goldstino corresponds to a local metric rescaling (the hole growing into the bulk).

\section{Conclusions}

The requirement of no new large contributions to flavour-changing neutral currents or CP violation
is one of the strongest constraints on the MSSM Lagrangian.
This article has investigated when supersymmetry breaking in string theory compactifications generates
flavour-universal soft terms. It has identified a set of sufficient conditions for flavour universality. These set of conditions
were called `mirror mediation'. They correspond to the factorisation of the hidden sector into two classes of field, with one
class sourcing flavour physics and the mirror class responsible for susy breaking.

This factorisation is artificial on the level of effective field theory, but naturally occurs within the effective
actions that apply in string compactifications.
In this case the sectors can be identified with the two mirror sectors of geometric moduli
associated with Calabi-Yau geometries, the K\"ahler and complex structure moduli.
At leading order these sectors are decoupled and do not mix.
In both IIA and IIB compactifications
the combination of holomorphy and shift symmetries implies that
the superpotential Yukawa couplings can depend only on one class of moduli.
In IIB compactifications flavour physics comes from
the complex structure moduli while for IIA models flavour physics originates with the K\"ahler moduli.
This was illustrated through the explicit formulae for the Yukawa couplings and matter field kinetic terms
in type II compactifications. The factorisation is broken at higher order in $g_s$ and $\alpha'$. This breaking is
suppressed at large volume and weak coupling.

Mirror mediation requires that supersymmetry breaking dominantly occurs in the mirror sector to that in which flavour physics
originates. In IIB models this implies that supersymmetry breaking should occur in the K\"ahler moduli sector. This is realised
by GKP flux compactifications, which have no-scale structure. In this case the
goldstino was shown to align exactly with the overall breathing mode. Going to the case of full moduli stabilisation,
we considered the large volume and KKLT models.
In the large volume case, the structure of supersymmetry breaking is largely
inherited from the GKP case. The goldstino aligns dominantly with the overall volume and susy breaking occurs in the
K\"ahler moduli sector. The large volume models therefore
give an explicit realisation of the full mirror mediation structure.
In KKLT-based models, the dominant susy breaking comes from the hidden sector used in uplifting
while susy breaking in the T-sector is subdominant. The structure of the soft terms depends on the details of the hidden sector
and is model-dependent.

There are several directions for future work. First, in any $\mc{N} =1 $ compactification
the factorisation of the moduli space will be broken at subleading order, corresponding to loop effects in either the
$\alpha'$ or $g_s$ expansions. Such breaking of factorisation, while suppressed by loop factors,
has the potential to lead to breaking of flavour universality in the soft terms.
The actual magnitude of corrections to universality have large phenomenological consequences,
and it would be very interesting to quantify this in specific models.
Secondly, the structure of mirror mediation suggests that in phenomenological
models of IIA string theory supersymmetry breaking should dominantly occur in the complex structure moduli sector, while
the K\"ahler moduli should be stabilised supersymmetrically.
It would be interesting to build fully stabilised models where this occurs, together with the generation of hierarchically low
supersymmetry breaking scales.

\section*{Acknowledgments}

I am funded by Trinity College, Cambridge. I thank Ben Allanach, Michele Cicoli, Kiwoon Choi, Thomas Grimm, Shamit Kachru, 
Fernando Quevedo, Kerim Suruliz and Scott Thomas for discussions and comments on the paper.

\appendix

\section{Intersection Angles in Calabi-Yau Backgrounds}
\label{appendixsec}

In toroidal models, the three angles that characterise D6-brane intersections are independent of flavour: repeated intersections
of the same two stacks occur with the same angles (however note that for fields of different gauge charges
the intersection angles vary). The purpose of the appendix is to investigate whether this still holds in
Calabi-Yau cases: we will find that for Calabi-Yau models the intersection angles are expected to be flavour non-universal,
even for fields of the same gauge charge.

The candidates for supersymmetric branes in IIA are D4, D6 and D8 branes. A Calabi-Yau has no 1- or 5-cycles, so this focuses our attention
on D6-branes.\footnote{although note supersymmetric coisotropic D8 branes can exist \cite{hepth0607219}.}
The Calabi-Yau breaks the $\mc{N}=8$ supersymmetry of type II string theory to
$\mc{N}=2$. A supersymmetric D-brane embedding further breaks this to $\mc{N}=1$.
Considering a single D6-brane, the condition that the embedding is supersymmetric is that the brane wraps a \emph{special Lagrangian} cycle.
A \emph{Lagrangian} 3-cycle $\Sigma$ is one for which the pullback of the K\"ahler form vanishes $J \vert_\Sigma = 0$, i.e.
\be
J({\bf e}_i, {\bf e}_j) = 0,
\ee
for all ${\bf e}_i, {\bf e}_j$ lying in the brane worldvolume $\Sigma$.
A \emph{special Lagrangian} 3-cycle is one which is volume-minimising, i.e. is
calibrated by the holomorphic (3,0) form $\Omega$,
\be
\rm{Im}(e^{i \theta} \Omega)\vert_{\Sigma} = 0, \qquad \rm{Re}(e^{i \theta} \Omega) \vert_{\Sigma} = \rm{vol}_{\Sigma},
\ee
for some phase $\theta$.
The brane breaks the $\mc{N}=2$ supersymmetry of the Calabi-Yau to $\mc{N}=1$. In terms of the $\mc{N}=2$ supercharges
$\psi_1$ and $\psi_2$, the
supercharge $\psi$ preserved by the D-brane is $\psi = \left( \cos \theta \right) \psi_1 +
\left( \sin \theta \right) \psi_2$.

D-branes carry positive charge ($+Q$) and tension ($+Q$).
To avoid a global RR tadpole while still preserving supersymmetry,
it is necessary to introduce objects that can simultaneously carry both negative tension and charge. Such objects are
orientifold O6-planes. These are located at fixed point sets of the orientifold action. They also wrap special
Lagrangian cycles and break the $\mc{N}=2$ supersymmetry to $\mc{N}=1$. In a globally $\mc{N}=1$ supersymmetric
configuration, all D-branes must preserve the same supercharge as the O-planes and so must
be calibrated with the same angle $\theta$.

As 3-dimensional objects in a 6-dimensional space, D6 branes
generically have pointlike intersections. If stacks of $N$ and $M$ branes intersect, then at the intersection locus
chiral matter is localised in the $(N, \bar{M})$ representation.
In intersecting brane worlds, the different gauge groups correspond to different brane stacks.
Family replication is due to nonvanishing
intersection numbers for distinct special Lagrangian 3-cycles. The number of chiral families is given by the topological
intersection number of distinct cycles. The physics of flavour is equivalent to the physics of the different intersections.

The intersection of two branes is always characterised
by three intersection angles. Each intersection is specified by angles $\theta_i$, $i = 1 \ldots 3$.
This is illustrated in figure \ref{Intersection}, which shows a brane embedded in local complex coordinates $z_1$, $z_2$, $z_3$ with
embedding $\rm{Im}(z_i) = 0$. It intersects with another brane embedded as $\hbox{Im}(e^{-i\theta_i} z_i) = 0$ at an intersection locus
$z_i = 0, i=1,2,3$ with intersection angles $\theta_1$, $\theta_2$ and $\theta_3$.
\FIGURE{\makebox[15cm]{\epsfxsize=12cm \epsfysize=5cm
\epsfbox{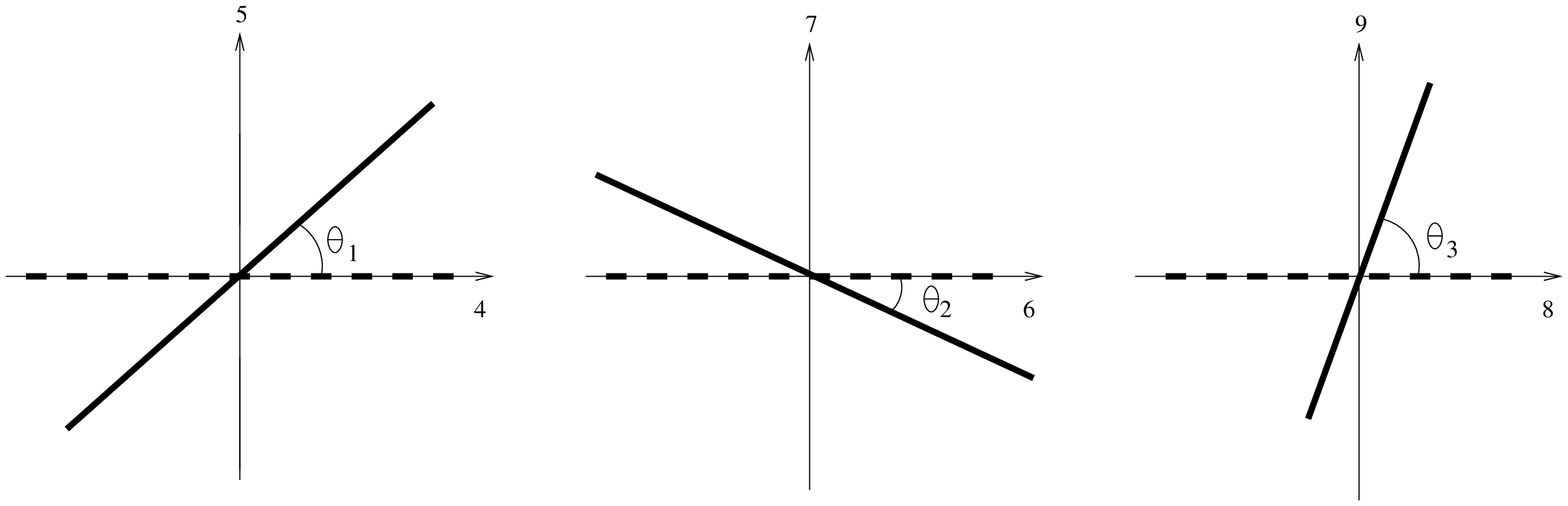}
}\caption{A locally factorisable brane intersection.}\label{Intersection}}

Any intersection is invariantly characterised by three angles. These angles are the
eigenvalues of the $SU(3)$ matrix that locally transforms one sLag into another. That is, if the first sLag corresponds
to $\rm{Re}(z_i) = 0$, and the second sLag corresponds to $\rm{Re}(w_i)=0$,
where now the intersection point $w_i = \mc{M}_{ij}z_j$
with $\mc{M} \in SU(3)$, the intersection angles $\theta_1$, $\theta_2$, $\theta_3$ are given by the eigenvalues of $\mc{M}_{ij}$.

For the case of sLags intersecting in a Calabi-Yau, we do not need to know the metric or the K\"ahler form
in order to compute the intersection angles. We here focus on a four-dimensional subspace, parametrised by
coordinates $x_1, x_2, x_3, x_4$, with $z_1 = x_1 + i x_2$ and
$z_2 = x_3 + i x_4$. We can write
\bea
g_{z_i z_j} & = & \frac{\partial x^\alpha}{\partial z^i} \frac{\partial x^\beta}{\partial z^j} g_{\alpha \beta} \nonumber \\
g_{z_i \bar{z}_j} & = & \frac{\partial x^\alpha}{\partial z^i} \frac{\partial x^\beta}{\partial \bar{z}^j} g_{\alpha \beta} \nonumber \\
g_{\bar{z}_i \bar{z}_j} & = & \frac{\partial x^\alpha}{\partial \bar{z}^i} \frac{\partial x^\beta}{\partial \bar{z}^j} g_{\alpha \beta}
\eea
In terms of the real metric $g_{\alpha \beta}$, we therefore have
\bea
g_{z_1 z_1} & = & \frac{1}{4}\left( g_{11} - g_{22} \right) - \frac{i}{2} g_{12}, \nonumber \\
g_{z_1 \bar{z}_1} & = & \frac{1}{4} \left( g_{11} + g_{22} \right), \nonumber \\
g_{\bar{z}_1 \bar{z}_1} & = & \frac{1}{4}\left( g_{11} - g_{22} \right) + \frac{i}{2} g_{12},
\eea
and likewise for $g_{z_2 z_2}$. We also have
\bea
g_{z_1 \bar{z}_2} & = & \frac{1}{4} \left( g_{13} + g_{24} \right) + \frac{i}{4} \left( g_{14} - g_{23} \right), \nonumber \\
g_{z_2 \bar{z}_1} & = & \frac{1}{4} \left( g_{13} + g_{24} \right) - \frac{i}{4} \left( g_{14} - g_{23} \right), \nonumber \\
g_{z_2 \bar{z}_1} & = & \left( g_{z_1 \bar{z}_2} \right)^*, \nonumber \\
g_{z_1 z_2} & = & \frac{1}{4} \left( g_{13} - g_{24} \right) - \frac{i}{4} \left( g_{14} + g_{23} \right) \nonumber \\
g_{\bar{z}_1 \bar{z}_2} & = & \frac{1}{4} \left( g_{13} - g_{24} \right) + \frac{i}{4} \left( g_{14} + g_{23} \right).
\eea
Fixing the complex structure is equivalent to requiring $g_{z_i z_j} = 0$, which imposes the conditions
$$
g_{11} = g_{22}, \qquad g_{33} = g_{44}, \qquad g_{12} = g_{34} = 0, \qquad g_{13} = g_{24}, \qquad g_{14} = - g_{23}.
$$
Any variations of the K\"ahler class which leaves the complex structure unaltered must satisfy the above conditions at the intersection
locus. We can write out the K\"ahler form as
\be
i J = g_{11} dx^1 \wedge dx^2 + g_{33} dx^3 \wedge dx^4 + g_{23} \left( dx^1 \wedge dx^3 + dx^2 \wedge dx^4 \right)
+ g_{13} \left( dx^1 \wedge dx^4 - dx^2 \wedge dx^3 \right)
\ee
This is the general local form of the K\"ahler class at $z=0$ that preserves the complex structure.

The fact that a brane wrapping the cycle $x_2 = x_4 = 0$ is special Lagrangian forces $g_{23} = 0$ along the worldvolume
of this brane, as otherwise the pull-back of the K\"ahler form would not vanish. As the intersecting brane is also Lagrangian,
we also require the pull-back of the K\"ahler form onto its worldvolume to vanish.
At the intersection locus, this implies that $g_{13} = 0$.

The fact that the two intersecting branes are wrapping Lagrangian cycles therefore restricts the locally non-vanishing
components of the metric to $g_{11} = g_{22}$ and $g_{33} = g_{44}$. From the geometry of the intersection, it is clear that
for branes wrapping Lagrangian cycles the local intersection angles can be determined independently of the
local value of the K\"ahler form at the intersection point.

Writing out $\Omega = (dx^1 + i dy^1) \wedge (dx^2 + i dy^2) \wedge (dx^3 + i dy^3)$,
it is easy to see that the special Lagrangian condition takes the well-known form
\be
\label{anglecondition}
\theta_1 + \theta_2 + \theta_3 = 0.
\ee

Unlike for D6-branes, IIB intersections are not pointlike and so it is not obvious that the chiral matter metrics can be characterised by
three angles. However mirror symmetry suggests that this is the case, and we assume here that the expression (\ref{anglecondition})
continues to hold for
IIB IBWs on a Calabi-Yau.

\subsection*{The Quintic}

The simplest Calabi-Yau is the Fermat quintic.
This is given by the hypersurface in $\mbb{P}^4$
\be
\sum z_i^5 = 0.
\ee
Special Lagrangian 3-cycles on the quintic have been studied in \cite{hepth9507158, hepth9906200, hepth0206038, hepth0210083}.
There exist a class of 625 sLag cycles, described by
$$
| k_1, k_2, k_3, k_4, k_5 \rangle = \{z_i :  \rm{Re}(\omega^{k_1} z_1) = \rm{Re}(\omega^{k_2} z_2) = \ldots = \rm{Re}(\omega^{k_5} z_5) = 0. \}
$$
Here $\omega^5 = 1$ and we may take $\omega = e^{2 \pi i/5}$. Due to the $\mbb{P}^4$ identification $(z_1, z_2, z_3, z_4, z_5) \equiv
\lambda(z_1, z_2, z_3, z_4, z_5)$, the cycles $|k_1, k_2, k_3, k_4, k_5 \rangle$ and $|k_1 + 1, k_2 + 1, k_3 + 1, k_4 + 1, k_5 + 1 \rangle$
are identified, and this family of sLags only contains 625 distinct examples.

The topological intersection matrix of these sLags was computed in \cite{hepth0210083}. The intersection number of the cyles
$|1,1,1,1,1 \rangle$ and $| k_1, k_2, k_3, k_4, k_5 \rangle$ is given by the coefficient of $g_1^{k_1} g_2^{k_2} g_3^{k_3} g_4^{k_4} g_5^{k_5}$
in
$$
\prod_{i=1}^5 (g_i + g_i^2 - g_i^3 - g_i^4) / \Big\{g_i^5 \equiv 1 \forall i, g_1 g_2 g_3 g_4 g_5 \equiv 1 \Big\}
$$
In general the cycle $| k_1, k_2, k_3, k_4, k_5>$ is calibrated by the form $\omega^{k_1 + k_2 + k_3 + k_4 + k_5} \Omega$ and so cycles are only
mutually sLag if $k_1 + k_2 + k_3 + k_4 + k_5 = 0 \rm{mod } 5$. There are then 5 sets of 125 mutually sLag - i.e. mutually supersymmetric - cycles.
It turns out that for these sets of cycles the intersection matrix vanishes: no
supersymmetric chiral matter can exist.

\subsection*{Other weighted $\mbb{P}^4$s}

The analysis for the quintic can be extended to other Calabi-Yaus that can be written
as a hypersurface in a weighted projective space.
These will in fact allow sLags to intersect with non-zero topological intersection numbers.

The weighted projective space $\mbb{P}^4_{[a,b,c,d,e]}$
is defined by
$$
(z_1, z_2, z_3, z_4, z_5) \sim (\lambda^a z_1, \lambda^b z_2, \lambda^c z_3, \lambda^d z_4, \lambda^e z_5).
$$
The degree $D=(a+b+c+d+e)$ hypersurface in $\mbb{P}^4_{[a,b,c,d,e]}$ satisfies the condition that
$c_1(\mc{M}) = 0$ \cite{Hubsch} and therefore admits a Calabi-Yau metric.
Up to non-polynomial deformations, the complex structure moduli space is described by the space
of inequivalent homogeneous polynomials of degree $D$. If $a,b,c,d,e$ all divide $D$, then the moduli space
contains a Fermat hypersurface. This hypersurface can be written as
\be
z_1^{n_1} + z_2^{n_2} + z_3^{n_3} + z_4^{n_4} + z_5^{n_5} = 0.
\ee
Here $n_1 = D/a, n_2 = D/b, \ldots n_5 = D/e$.
At the Fermat locus an enhanced $\mbb{Z}^{n_2} \otimes \mbb{Z}^{n_3} \otimes \mbb{Z}^{n_4} \otimes \mbb{Z}^{n_5}$ discrete symmetry
exists, corresponding to coordinate rotations.
Special Lagrangian cycles can be constructed for these Calabi-Yaus in analogy to the
construction for the quintic.
The Fermat hypersurface admits an antiholomorphic involution $z \to \bar{z}$. This involution has as its fixed
point the fundamental sLag, which we denote by $|0,0,0,0,0>$. This sLag is given by
$$
(x_1, x_2, x_3, x_4, x_5) \hbox{ with } x_1^{D/a} + x_2^{D/b} + x_3^{D/c} + x_4^{D/d} + x_5^{D/e} = 0.
$$

We can construct a family of sLags by acting with the discrete symmetry generators on the fundamental sLag.
We denote by $|\lambda_1, \lambda_2, \lambda_3, \lambda_4, \lambda_5>$ the sLag that arises by rotating the
fundamental sLag by the following discrete transformation
\bea
z_1 & \to & e^{2 \pi i \lambda_1 a/D} z_1, \nonumber \\
z_2 & \to & e^{2 \pi i \lambda_2 b/D} z_2, \nonumber \\
z_3 & \to & e^{2 \pi i \lambda_3 c/D} z_3, \nonumber \\
z_4 & \to & e^{2 \pi i \lambda_4 d/D} z_4, \nonumber \\
z_5 & \to & e^{2 \pi i \lambda_5 e/D} z_5.
\label{rot}
\eea
In cases where some of the $a, \ldots, e$ are even, further families of sLags not included in (\ref{rot}) may be found
using half-integral
actions of the discrete symmetry generators; for example, when $n_1$ is even this corresponds to the fixed point set
under the involution
\bea
z_1 & \to & \exp{2 \pi i/n_1} z_1, \nonumber \\
z_2 & \to & \bar{z}_2, \nonumber \\
z_5 & \to & \bar{z}_5. \nonumber
\eea
This is familiar from IIA toroidal orientifolds, where for even orientifold actions
there exist distinct O-planes that are not related by the orbifold action.
Using the above procedure several families of sLags may be generated for
the weighted projective spaces. As the cycles are known explicitly, the intersection numbers may be computed
using geometric means, by explicitly locating the intersection points. Through examination of the
local intersection it is possible to compute the topological intersection number, counting the sign of
each intersection.

Mutually supersymmetric sLags must be calibrated by the same holomorphic 3-form. From the definition of $\Omega$,
$$
\Omega = \frac{1}{2 \pi i} \int \frac{z_5 dz_1 \wedge dz_2 \wedge dz_3 \wedge dz_4}{p(z)},
$$
it follows that $|\lambda_1, \lambda_2, \lambda_3, \lambda_4, \lambda_5>$ is calibrated by the same form as
$|0, 0, 0, 0, 0>$ as long as $a \lambda_1 + b \lambda_2 + c \lambda_3 + d \lambda_4 + e \lambda_5 = 0 \textrm{ mod } D$.

Using the above construction of sLags for general weighted projective spaces, we can investigate the
intersection properties of sLags beyond the case of the quintic.
We have considered all weighted projective spaces where $a, \ldots, e$ are all odd. These are
$$
\mbb{P}^4_{[1,1,1,1,1]}, \quad \mbb{P}^4_{[1,1,1,3,3]}, \quad \mbb{P}^4_{[1,3,3,7,7]},
\quad \mbb{P}^4_{[1,1,3,5,5]}, \quad \mbb{P}^4_{[1,3,3,3,5]}, \quad \mbb{P}^4_{[1,5,9,15,15]}.
$$
From this we can conclude that for generic mutually supersymmetric
sLags with multiple chiral intersections the intersection angles differ
between intersections.

As an example, we consider the Calabi-Yau hypersurface
in $\mbb{P}^4_{[1,3,3,3,5]}$. This space is defined by
$$
(z_1, z_2, z_3, z_4, z_5) \sim (z_1, \lambda^3 z_2, \lambda^3 z_3, \lambda^3 z_4, \lambda^5 z_5).
$$
The Fermat hypersurface is given by
\be
z_1^{15} + z_2^5 + z_3^5 + z_4^5 + z_5^3 = 0,
\ee
and admits a discrete $\mbb{Z}^5 \otimes \mbb{Z}^5 \otimes \mbb{Z}^5 \otimes \mbb{Z}^3$ symmetry. The fundamental
involution is $z_i \to \bar{z}_i$ and has a fixed point set $(\lambda x_1, \lambda^3 x_2, \lambda^3 x_3, \lambda^3 x_4,
\lambda^5 x_5)$ with $x \in \mbb{R}$ and
$$
x_1^{15} + x_2^5 + x_3^5 + x_4^5 + x_5^3 = 0.
$$
This has a unique solution for $x_5$ given $x_1, \ldots x_4$ and the fixed point set is topologically an
$\mbb{R} \mbb{P}^3$. We denote this fundamental sLag by $|0,0,0,0,0>$.

Now consider the sLag $|0,1,1,3,0>$. This is obtained by acting on the fundamental sLag with
$$
z_1 \to z_1, z_2 \to e^{2 \pi i/5} z_2, z_3 \to e^{2 \pi i/5} z_3, z_4 \to e^{6 \pi i/5} z_4, z_5 \to z_5
$$
It has fixed point set
$$
(\lambda x_1, \lambda^3 e^{2 \pi i/5} x_2, \lambda^3 e^{2 \pi i/5} x_3, \lambda^3 e^{6 \pi i/5} x_4, \lambda^5 x_5)
$$
with $x_1^{15} + x_2^5 + x_3^5 + x_4^5 +x_5^3 = 0.$

The sLags $|0,0,0,0,0>$ and $|0,1,1,3,0>$ intersect at the following loci:
$$
(1,0,0,0,-1)
$$
and
$$
(0,0,0,1,-1) \equiv (0,0,0,e^{6 \pi i/5}, -1).
$$
For the first intersection, we can go to local complex coordinates $(w_1, w_2, w_3)$ defined by
$$
(1, w_1, w_2, w_3, (-1 +w_1^5 + w_2^5 + w_3^5)^{1/3}).
$$
In these coordinates the sLag $|0,0,0,0,0>$ is described by
$$
\hbox{Re}(w_1) = \hbox{Re}(w_2) = \hbox{Re}(w_3) = 0
$$
while the sLag $|0,1,1,3,0>$ is described by
$$
\hbox{Re}(e^{-2 \pi i/5}w_1) = \hbox{Re}(e^{-2 \pi i/5}w_2) = \hbox{Re}(e^{-6 \pi i/5}w_3).
$$
From this we can deduce the local intersection angles at (1,0,0,0,-1) to be $(2 \pi/5, 2 \pi / 5, 6 \pi / 5)$.

For the second intersection, we can define local complex coordinates $(w_1, w_2, w_3)$ by
$$
(w_1, w_2, w_3, 1, (-1 + w_1^{15} + w_2^{5} + w_3^5)^{1/3}).
$$
In these coordinates the sLag $|0,0,0,0,0>$ is described by
$$
\hbox{Re}(w_1) = \hbox{Re}(w_2) = \hbox{Re}(w_3) = 0
$$
and the sLag $|0,1,1,3,0>$ is described by
$$
\hbox{Re}(e^{2 \pi i/5}w_1) = \hbox{Re}(e^{4 \pi i/5}w_2) = \hbox{Re}(e^{4 \pi i/5}w_3).
$$
From this we can deduce the local intersection angles to be $(-2 \pi/5, -4 \pi / 5, -4 \pi / 5)$.

The sLags $|0,0,0,0,0>$ and $|0,1,1,3,0>$ also intersect along the $S^1$ $(0, x_2, x_3, 0, x_5)$.
The normal bundle of a sLag is equivalent to its tangent bundle. As the tangent bundle of an $S^1$ admits a global
non-vanishing section, there likewise admits a global non-vanishing section of the normal bundle, and so this intersection
is not topological; it can be removed by a deformation of the cycles.

The sign of each intersection is given by $\prod_i \rm{sgn}(e^{i \theta_i})$, where $\theta_i$ are the three intersection angles.
The sign of each intersection above is $-1$ and so the topological intersection number of the above two cycles is $-2$.
This shows that for Calabi-Yau interesecting brane worlds, there does not seem to be a reason for
the intersection angles of two brane stacks to be family-universal: fields of the same charges but different families can have
different intersection angles.

Similar behaviour can be seen for sLags in $\mbb{P}^4_{[1,1,1,3,3]}$ and $\mbb{P}^4_{[1,1,3,5,5]}$. In the former
case, the intersection of $|0,0,0,0,0>$ and $|0,0,3,1,1>$ gives family non-universal intersection angles, whereas in the
latter cases non-universality is found for the intersections of $|0,0,0,0,0>$ with the sLags
$|0,1,3,1,0>$, $|0,2,1,2,0>$ and $|0,4,2,1,0>$.

\end{document}